# Automatic Calculation of the Transition Temperatures for two-dimensional Heisenberg type Magnets

Haichang Lu[1,2,3]*, Tai Yang[1], Zhimei Sun[4], John Robertson[3]* and Weisheng Zhao[1,2]*

[1]*Fert Beijing Institute, MIIT Key Laboratory of Spintronics, School of Integrated Circuit Science and Engineering, Beihang University, Beijing, 100191, China2*

[2]*National Key Lab of Spintronics, Institute of International Innovation, Beihang University, Yuhang District, Hanzhou, 311115, China*

[3]*Engineering Department, Cambridge University, Cambridge CB2 1PZ, UK*

[4]*School of Materials Science and Engineering, Beihang University, Beijing 100191, China*

Email: HaichangLu@buaa.edu.cn
jr214@cam.ac.uk
weisheng.zhao@buaa.edu.cn

Theoretical prediction of the $2^{nd}$-order magnetic transition temperature ($T_M$) used to be arduous. Here, we develop a first principle-based, fully automatic structure-to-$T_M$ method for two-dimensional (2D) magnets whose effective Hamiltonians follow the Heisenberg model. The Heisenberg exchanges, which can be calculated to an arbitrary shell, are transferred into the Monte Carlo calculation. Using Cr-based magnets as the showcases, we show that our method is a powerful tool to study the 2D magnets in two aspects. First, considering long-range exchanges enables us to identify the spin frustration in the suspended $CrTe_2$ monolayer, whereas the heterostructure calculations reveal that the ferromagnetism can be recovered if the monolayer $CrTe_2$ is grown onto various 2D substrates. Second, we realize a high-throughput screening of novel magnets discovered by random structure searches. Six 2D Cr chalcogenides are selected to have high $T_M$. Our work provides a new insight for the study of 2D magnets and helps accelerate the pace of magnetic materials data-mining.

## I. INTRODUCTION

Magnetic materials have wide applications in data storage and spintronic devices, presenting a promising platform for future low-energy consumption technology [1,2]. Among all the important features of magnets, high transition temperature ($T_M$) is the prerequisite to ensure the non-volatile feature of the devices at the working temperature. Understanding the origin of magnetism is essential for material selection, especially for the emergent two-dimensional (2D) magnets which spark huge attention as promising candidates for next-generation magnetic memory. As most of the materials discovered cannot sustain magnetic orders above room temperature [3,4], many materials remain to be explored.

Among the earliest discoveries of intrinsic magnetism in 2D van der Waals materials, many of them are Ising-type or Heisenberg-type Cr and Mn-based magnets, which are either semiconductive or metallic, such as CrI$_3$ [5], CrGeTe$_3$ [6], CrTe$_2$ [7,8], MnSe$_2$ [9], MnPS$_3$ [10]. Theoretical prediction of T$_M$ involves calculating the magnetic properties like the exchange interactions, and magnetic anisotropy, which are essential to understand the origin of the magnetism and design novel materials. The bilinear Heisenberg exchanges (HEs) are often the most important energy term in the Hamiltonian to determine T$_M$. Extraction of the HEs is a challenging task, which has been done by various methods [11]. For itinerant magnets, the magnetic force method based on linear response theory is widely adopted [12-15]. Whereas for magnets with localized electrons, HEs can be obtained from indirect calculation known as energy mapping analysis [16], where the effective spin Hamiltonian is under the framework of the Heisenberg model [17]. It can be further extended to itinerant magnets by mapping into the Heisenberg model with Ruderman-Kittel exchange [18], especially in the presence of anisotropic exchange interactions [19,20]. The energy mapping method based on density functional theory (DFT) is prevalent because it is a straightforward technique, only requiring the calculation of total energies for different spin configurations.

To date, many works of predicting T$_M$ are done involving self-explore analysis of certain materials, considering only a few nearest neighbor HEs, which are short-range [21-25]. Here, we aim to improve the calculation procedure in two aspects. First, long-range HEs can be non-negligible, we design a method to obtain a series of the HEs to an arbitrary order of shell for either ferromagnets or antiferromagnets which possess Heisenberg-type spin Hamiltonian and various stable spin configurations. Second, all processes for HEs calculation can be done automatically by computer. Previously, high-throughput Curie temperature (T$_C$) calculation for 2D ferromagnets has been reported [26], considering HEs to the 4$^{th}$ shell. In this work, we present the HEs of various magnets to the 4$^{th}$ ~16$^{th}$ shell.

We develop an end-to-end code called Automated Structure to Temperature-Dependent Magnetism (*ASTDM*), which realizes high-throughput prediction of T$_M$ for any kind of spin-lattice. We adopt the energy mapping method to obtain the HEs. There are many kinds of energy mapping methods. For example, the four-state method treats the $i, j$ sites like defects and requires manually freezing the spin directions [16]. It has the advantage of extracting anisotropic exchange by delicate spin manipulations. However, it inevitably introduces mirror image errors. A large supercell is needed to reduce the error, especially for long-range HEs. One can also do a spin rotation to fit the energy with the rotation angle, it can be hard to maintain the spin-frozen configuration using DFT. Here, we propose a method to extract HEs by enumerating plausible antiferromagnetic (AFM) states and solving the spin-dependent Heisenberg matrix. We find that various AFM states are often easy to maintain in Cr and Mn-based magnets. Similarly, a high-throughput workflow for calculating the ground state of any magnets was proposed recently by listing metastable states [27]. Our method is not only free of the mirror image error but also computationally efficient by the one-off output of a series of HEs. For n number of HEs, only n+1 total energy calculations are needed. Aside from HEs, we also consider the onsite single-ion magnetic anisotropy energy (MAE) which is important in sustaining the magnetic order in 2D magnets. We then forward them to the recently introduced semiclassical Metropolis Monte Carlo (SMC) simulation [28] to obtain magnetic properties helping identify $T_M$ such as magnetic moment, heat capacity, and magnetic susceptibility. Our method is user-friendly for high-throughput calculation, requiring no more than an input structure with a few calculation parameters (see details in the supplementary material (SM) III).

We first test the code with various experimental known magnets. Then we focus on the 1T chromium ditelluride ($CrTe_2$), one of the experimental known ferromagnets to have $T_C$ approaching room temperature. Down to the monolayer limit, spin frustration exists due to large long-range HEs and small short-range HEs. The Ferromagnetic (FM) order of the ML $CrTe_2$ can be recovered by van der Waals contact with various 2D substrates. Finally, we perform a random structure search of Cr binary compounds to screen 2D magnets with high transition temperatures. We identified six magnets. The 1T' $CrS_2$, 1T1T $Cr_2S_3$, 1T2H $Cr_2Te_3$ bulk and the 1T2H of $Cr_3Te_4$ are FM. The 1T2H $Cr_2X_3$ (X=S, Se, Te (ML)) are AFM. Their $T_M$ are calculated to be above room temperature.

## II. *ASTDM* WORKFLOW

The Heisenberg Hamiltonian can be written as

$$H_{mag} = E_0 + \sum_{<ij>} J_{ij}\vec{s}_i \cdot \vec{s}_j + \sum_i A_{\vec{u}}(\vec{s}_i \cdot \vec{u})^2$$

where $E_0$ is the non-magnetic energy, $J_{ij}$ is the Heisenberg exchange, $\vec{s}_i/\vec{s}_j$ is the spin vector on the $i/j$ spin site. The second term is the single ion magnetic anisotropy energy, which originates from the structure anisotropy and the spin-orbit coupling (SOC) effect, $\vec{u}$ is the unit vector pointing to the anisotropic axis. To predict the transition temperature, both the HEs and the MAE should be calculated.

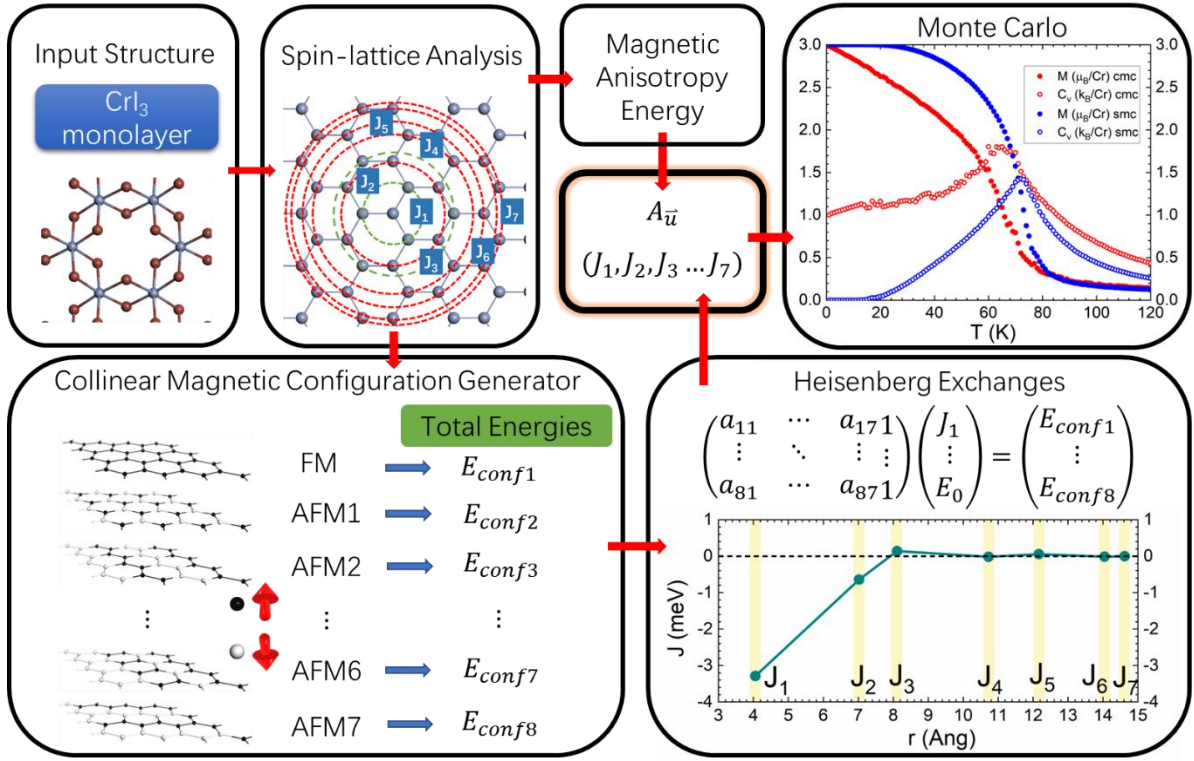

FIG. 1. Workflow of *ASTDM*, using two-dimensional CrI$_3$ as an example. [Input structure]: Cr atoms are marked in grey and I atoms are marked in brown. [Spin-lattice analysis]: for a spin site Cr, HEs with three neighbors are marked in green circles, and HEs with six neighbors are marked in red circles. [Collinear magnetic configuration generator]: black points are set spin-up, and white points are set spin-down. [Monte Carlo]: dots are magnetic moments M, hollow dots are heat capacity C$_V$, which helps identify the transition temperature. Red is for CMC and blue is for SMC.

The first step is to analyze the structure and determine the number of HEs that need to be calculated. Fig. 1 demonstrates the workflow using the monolayer (ML) chromium triiodide (CrI$_3$). The input structure should be geometry-optimized and the size of the spin moment in each ion is also written, see more details in **supplementary information (SI) III**. First, magnetic ions are identified to construct the spin-lattice and classified by their species and spin sizes. Then HEs are categorized by the two connecting magnetic ions $i$ and $j$ and the distance $r_{ij}$. Long-range HEs are often less important since HEs are related to distance. Therefore, only the $J_{ij}$ satisfying $r_{ij} = |\vec{r}_i - \vec{r}_j| < r_c$ (cutoff distance) are counted. Test of $r_c$ is performed for each material investigated to make sure no important HEs are neglected. Fig. 1 [Spin-lattice Analysis] shows the spin-lattice and the naming of HEs for the ML CrI$_3$. CrI$_3$ is a two-dimensional semiconductor with a hexagonal lattice and FM ground state. The $r_c$ is set to be 15Å, so that HEs are considered to the 7$^{th}$ shell. *ASTDM* will generate *seven* meta-stable AFM configurations plus the FM configuration. We sort the HEs ($J_1, J_2, J_3 ...$) by $r = |\boldsymbol{r}_i - \boldsymbol{r}_j|$. The total energy of each configuration can be written as:

$$E_{conf\ m} = E_0 + \sum_j a_{mn} J_n$$

where $m = 1$ is for FM, and $m = 2\sim 8$ are for AFM1~AFM7, shown in Fig. 1 [Collinear Magnetic Configuration Generator]. The matrix element $a_{mn}$ is calculated by summing up the pairwise exchange energies related to $J_n$ for the m$^{th}$ spin configuration. The spin dimer energy for site $i$ and $j$ with the exchange interaction $J_n$ can be written as $J_n s_i s_j$ for if the spins are parallel, $-J_n s_i s_j$ for if the spins are antiparallel. The energy of each configuration will be calculated by density functional theory (DFT). By solving the $8 \times 8$ matrix equation, $(J_1, J_2, J_3 \ldots J_7)$ and $E_0$ can be deduced. Noted the matrix should be non-singular, so $\vec{a}_i = (a_{i1}, a_{i2}, \ldots, a_{i7}, 1)$ should be linearly independent of any other $\vec{a}_j$. The recursive function is used to iterate through every possible AFM configuration from the original input cell to expanded supercells, prioritized by the surface area to minimize the computational cost. Any current generated configuration not orthogonal to the previously accepted configurations is discarded. Supercells are expanded doubly each time in each direction. Once *eight* orthogonal configurations are found, the process is terminated. The size of the supercell is set to be the minimum size to interpret every configuration.

Next, input files for the first-principle total energy calculations are generated by the input parameters listed in SM III. Details of the calculation parameters in this work are illustrated in SM I. After the calculations are finished, *ASTDM* picks up the total energies from the result file. Apart from energy collection, it will also examine whether the output magnetic configurations are consistent with the initial setups. If all configurations pass the check, the HEs are obtained via Gaussian elimination of the matrix. Fig. 1 [Heisenberg Exchanges] shows $(J_1, J_2, J_3 \ldots J_7)$ of ML CrI$_3$. In this case, $J_1, J_2$ are important to determine T$_C$, so $r_c = 9$Å is enough. Nevertheless, it is good to know how HE decade from the 1$^{st}$ shell to the 8$^{th}$ shell.

Apart from the HEs, the onsite MAE is also calculated by

$$A_{\vec{u}} = \frac{E_{axis} - E_{plane}}{s^2}$$

where $E_{axis}$ and $E_{plane}$ are the total energy per Cr of the FM configurations with all spins parallel and perpendicular to $\vec{u}$, respectively. For CrI$_3$, we find an easy axis along the z-direction, revealing perpendicular magnetic anisotropy (PMA). The MAE is -0.025meV/Cr.

Finally, the code moves on to the Monte Carlo calculation to calculate the temperature-dependent thermal properties such as the average magnetic moment M, the heat capacity C$_V$, the magnetic susceptibility $\chi$, and the normalized entropy S (see SM II and Fig. S2 for more details). *ASTDM* is capable of doing either classical Monte Carlo (CMC) or semiclassical Monte Carlo (SMC). The magnetization and the heat capacity are shown in Fig. 1 [Monte Carlo]. SMC partially captures the nature of quantum spins by recently introduced local quantization[30]. Note that all HEs should be scaled to quantum Heisenberg exchange by $J^Q = J^{Cl} \frac{S}{S+1}$. The MC calculation is proceeded to finite size test (Fig. S3) and the loop test (Fig. S4) for each magnet investigated in this work. Finally, we can visualize the spin textures at a given temperature to help identify the magnetic state.

## III. RESULTS

### A. Material investigations

We test the transition temperatures of several experimental known 2D magnets (both the monolayer and the bulk forms shown in Table S1, Fig. S6~S14) and conventional 3D magnets (Table S2, Fig. S15~S21). We perform $r_c$ and $k_c$ (see the definition of $k_c$ in SM III) test for each material, examples are shown in Fig. S1. The trend of HEs is studied. Fig. S6~S21 show the crystal structure, spatial distance vs HEs ($J(r)$), and the magnetization (M), heat capacity ($C_V$) using the fluctuation-dissipation theorem. It is worthwhile noticing that the 2D spin lattice with in-plane anisotropy has no long-range order, which is like the XY model with BKT (Berezinskii, Kosterlitz, and Thouless) phase transition. We add an external in-plane magnetic field to create an effective easy axis anisotropy in the plane, which recovers magnetic order, shown in Fig. S5. By looking at the $C_V$, we find that a field up to 1T contributes little change to the $T_M$.

Semiconductive magnets are well described by the Heisenberg model, as they possess localized moments. The 2D magnets we inspect here have been tested that the bilinear Heisenberg exchanges are the major coefficients determining $T_M$ by confirming that the spin energy vs. spin rotation angle roughly follows $E(\theta) \propto cos\theta$ [29]. We only consider the Heisenberg exchange as other exchanges are often an order smaller and usually contribute less to the $T_M$. We find the intralayer HEs are similar in both ML and bulk form, so they are not thickness-dependent. As interlayer HEs are involved, $T_M$ in bulk form is often larger than that in ML form.

The Heisenberg model is not primarily designed for itinerant magnets. However, if the unpaired electrons are localized and the interactions between the localized spins are via conduction electrons, the itinerant system can be interpreted by the extended Heisenberg model with the trend of HEs following Ruderman-Kittel-Kasuya-Yosida (RKKY) dispersion [18]. Here we study Cr chalcogenides. Though being itinerant magnets, they possess various AFM metastable states without spin quench, in contrast to many other bulk itinerant magnets such as Fe, Co, and Ni where AFM states are not stable, and Cr where FM state cannot be obtained. We also calculate the spin-resolved partial density of states (PDOS) in Fig. S1(a)(d). The spin-up Cr d bands are localized 1eV beneath the Fermi level ($E_F$), meaning that the level of localization is good for RKKY. By performing the $k_c$ test (Fig. S1 (b)(c)(e)(f)), we find that the HEs are more distance-dependent rather than insulators whose HEs also depend on spin-lattice. The $J(r)$ roughly follows the RKKY curve, whose 2D form is [30,31]:

$$J(r) \sim -\frac{\cos(2k_F r)}{(2k_F r)^2}$$

and 3D form is [32]:

$$J(r) \sim -\frac{\sin(2k_F r) - (2k_F r)\cos(2k_F r)}{(k_F r)^4}$$

$k_F$ is the Fermi vector. We study CrTe$_2$, the HEs, and MC results are shown in Fig. S14 and Fig. 2. We fit the $J(r)$ of ML and bulk to the 2D and 3D RKKY, respectively. $k_F$ of ML is larger than that of bulk, as the quantum confinement effect is stronger as the dimension shrinks and d electrons become more delocalized, as shown in Fig. S1(d). PDOS at the Fermi level $D(E_F)$ is significantly enhanced from bulk. One would naively conclude dominant HEs determining the magnetism should be short-range HEs whereas long-range HEs should be negligible because the exchange is related to the spatial distance between the exchange sites. This is true for insulators. However, electrons are delocalized in itinerant magnets, so the wavefunction overlap is important. Therefore, a larger $r_c$ is required. We find $J_1$ is overridden $J_2$ and $J_4 > 0$, which causes spin frustration. We examine the spin texture in the presence of the in-plane field of 1T to exclude the effect of planar anisotropy and find spin frustration is responsible for the disappearance of magnetization. For ML, T$_M$=255K. The spin frustration vanishes for more than 2ML. As $J_{z0} > 0$, the ground state is A-type AFM, with T$_N$ close to 300K. Approaching bulk, we obtain a T$_C$ of 395K. We also test with Hubbard U correction in Fig. S1, both $D(E_F)$ and $k_F$ are reduced due to stronger electron Coulomb repulsion.

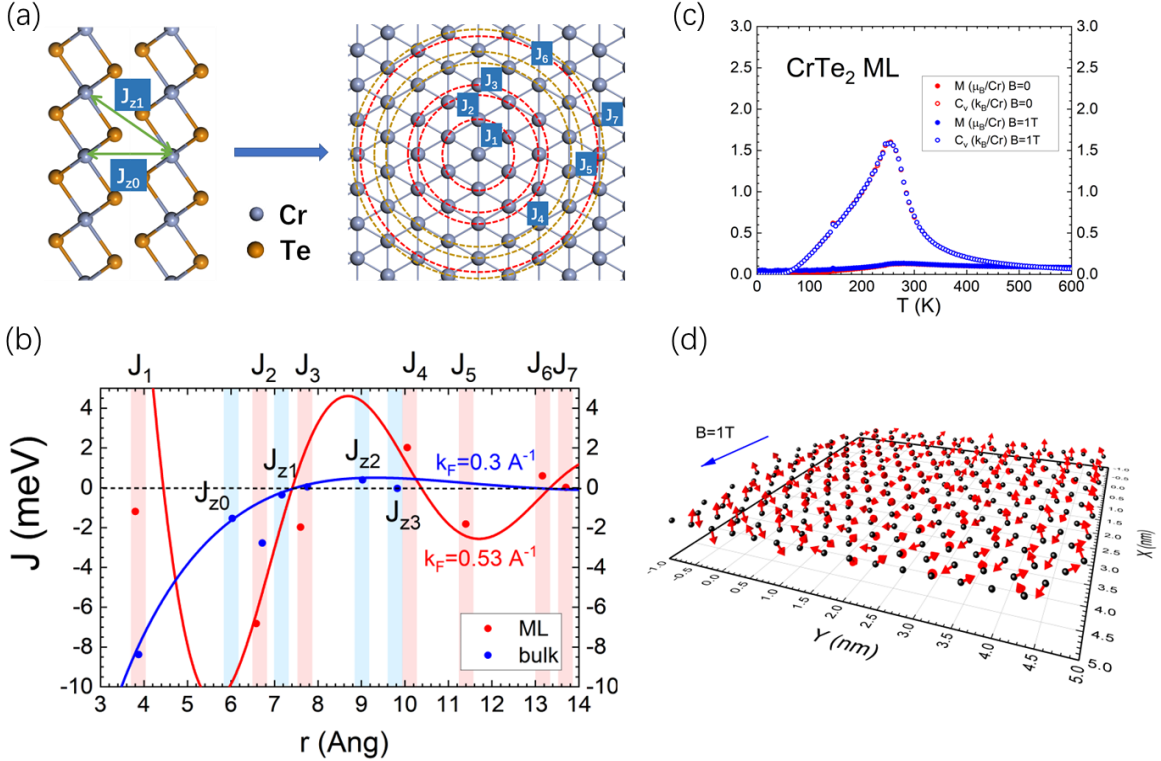

FIG. 2. (a) Atomic structure and spin-lattice of $CrTe_2$ with sorted HEs. The intralayer HEs and interlayer HEs are sorted separately by distance. (b) RKKY fitting of ML and bulk $CrTe_2$. (c) M and $C_V$ for ML $CrTe_2$, zero-field, and 1T in-plane field. (d) Spin frustration with the presence of an in-plane field.

### B. Interfacial effect

*ASTDM* is also capable of extracting the HEs in monolayer magnets contacting with substrates. The interfacial engineering is an interesting subject which aims to alter the electronic structure near the interface. It is particularly popular for 2D heterostructures. Down to ultra-thin layers, the interfacial effect is not negligible and often overcomes the bulk effect. Therefore, we investigate the interfacial effect on the HEs.

Recently, we also demonstrated how the substrate can dramatically change the HEs to revise the Curie temperature [33]. The dangling bonds at the interface of the 3D substrate contribute to the d-band shifting of the magnets. In this work, we focus on the van der Waals contact without dangling bond. Note that the categorizing of HEs is done by identifying the spin pairs and the distance. Therefore, it is important to make sure the substrate does not deform the magnet, which is true in our cases as the 2D magnets are physisorbed onto substrates. The substrate also breaks the symmetry in the z direction, but we only consider one layer of the magnet and one layer of the substrate. We consider the SOC effect for the substrates with heavy transition metal.

We study the heterostructure consisting of metallic ML $CrTe_2$. We select several 2D substrates to construct the van der Waals heterostructures, in which the lattice mismatches are less than 3%. Table 1 shows how the heterostructures are constructed. Observing Fig. 3 (b), we find that how HEs are reshaped is mainly affected by the type of the lattice match. For example, $CrTe_2$ is 1:1 matched to $PtSe_2$, so HEs do not change much. It is 2:3 matched to graphene and hexagonal boron nitride (h-BN), reshaping HEs in the same way. This is also true for transition metal dichalcogenides (TMDs) $MoSe_2$ and $WSe_2$. For $Bi_2Te_3$, HEs are greatly reversed. Apart from $PtSe_2$, spin frustration is eliminated. SOC effect is stronger in $Bi_2Te_3$ than in $PtSe_2$ and $WSe_2$, which is also seen in magnetic insulator $CrI_3$ (Fig. S22). The effect in the van der Waals interface should be an order less than interfaces with dangling bonds, so no dramatic change of the Curie temperature is expected, but it reshapes HEs in a way of symmetry breaking, so the contact configuration is important. Our calculation provides insights into why single-layer ferromagnetism is observed in experiments [7,8,34].

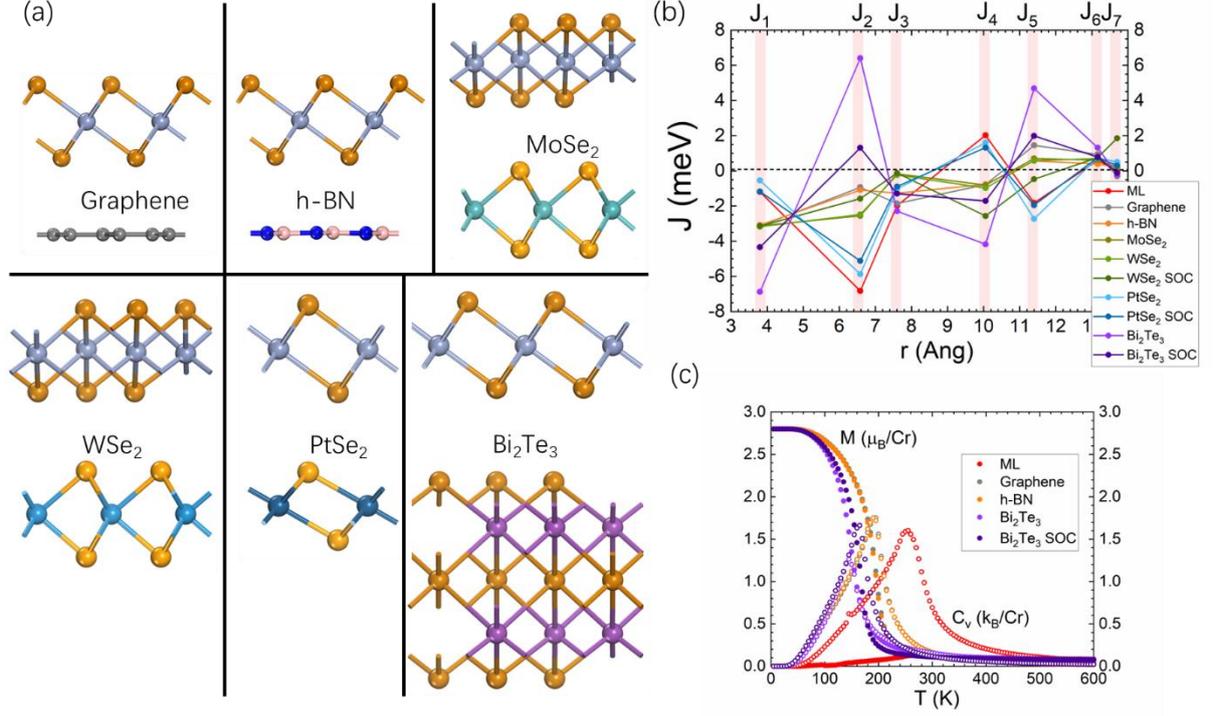

FIG. 3. (a) Atomic structure, (b) HEs, and (c) M, $C_V$ f of the $CrTe_2$-substrate heterostructure.

TABLE I. Comparison of $CrTe_2$ ML contacting with different substrates, Lattice match, mismatch, Curie temperature, and MAE.

| Substrate | SOC | Lattice match | Mismatch (%) | $T_C$ (K) | MAE ($\mu$eV/Cr) |
|---|---|---|---|---|---|
| Suspended | × | / | / | 255 | 90.99 |
| Graphene | × | 2:3 | 2.88 | 195 | 62.75 |
| h-BN | × | 2:3 | 1.14 | 190 | 110.19 |
| $MoSe_2$ | × | $\sqrt{3}$:2 | -0.91 | 185 | 238.71 |
| $WSe_2$ | × | $\sqrt{3}$:2 | -0.91 | 190 | 179.54 |
|  | √ |  |  | 185 |  |
| $PtSe_2$ | × | 1:1 | 1.59 | 295 | 389.45 |
|  | √ |  |  | 210 |  |
| $Bi_2Te_3$ | × | 2:$\sqrt{3}$ | -0.75 | 145 | 90.03 |
|  | √ |  |  | 165 |  |

SOC: × for no SOC effect is considered, √ for considering SOC effect.

Lattice match: $CrTe_2$ supercell to the substrate supercell.

## C. Structure search

Finally, we use *ASTDM* to predict the $T_M$ of materials that have not been experimentally realized. To find 2D magnets with potential high $T_M$, one would require the metalloid with weak electronegativity. Apart from the metalloid, spin-lattice with denser sites is preferred to have more HEs. Therefore, one would seek materials where the magnetic ion ratio is high. However, materials satisfying these might not exist, or the formation energies are too low to be fabricated experimentally. In light of this, we combine random structure searches and *ASTDM* to seek energetically favored structures with high $T_M$.

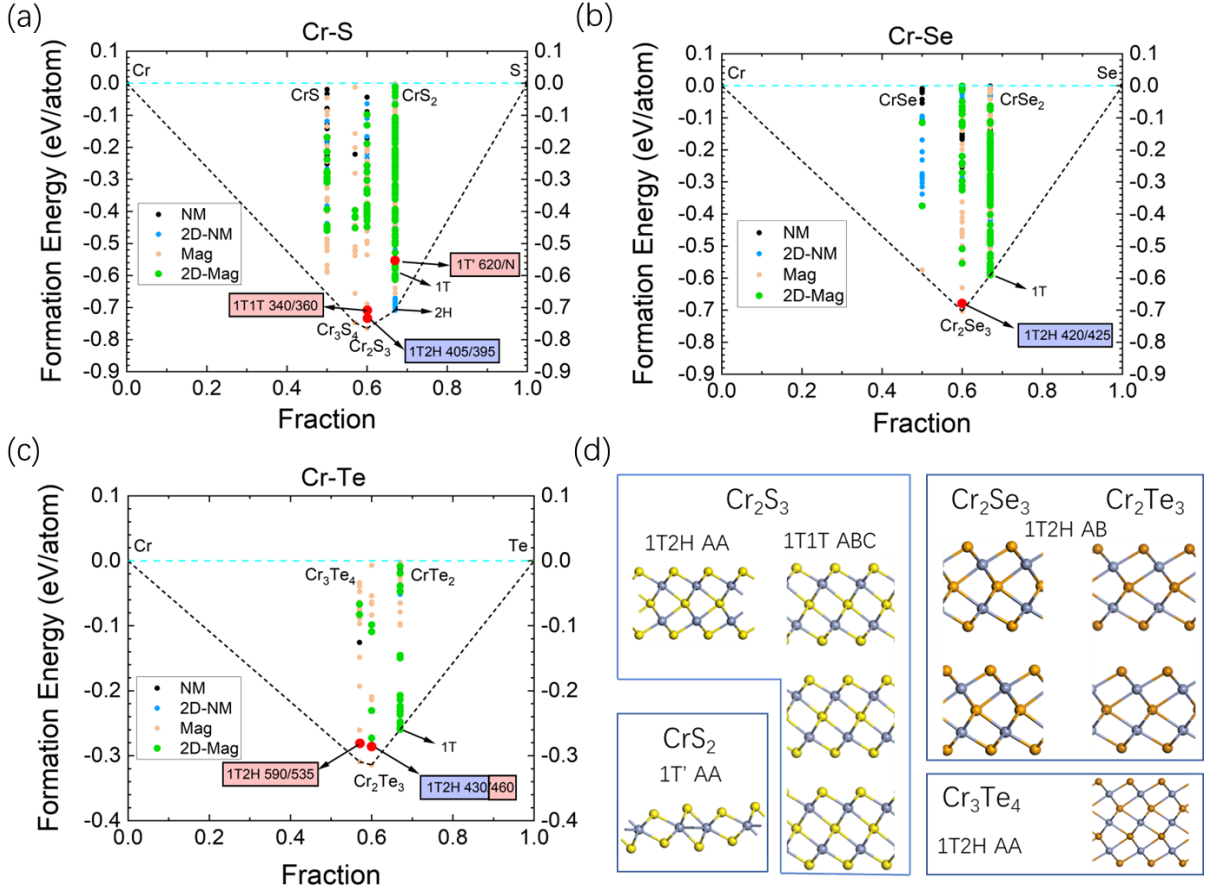

FIG. 4. Convex hull of (a) Cr-S, (b) Cr-Se, (c) Cr-Te binary compounds. Selected magnets marked with structure type. Transition temperatures are written by ML/bulk, in red for FM magnets, and blue for AFM magnets. (d) Selected structures with $T_M$ approaching or higher than room temperature.

We focus on Cr-based binary magnets. The metalloid is chosen to be S, Se, Te. We implement the ab initio random structure search (*AIRSS*) code [35] which is integrated with *ab initio* code CASTEP [36]. Fig. 4 shows the convex hull of the Cr-S, Cr-Se, and Cr-Te binary compounds. All structures with negative formation energies are plotted and classified by 2D/3D magnets/non-magnets. All 2D materials are screened using the rank determination algorithm (RDA) method [37]. Then we pick six 2D magnets whose 'energies above the hull' are reasonably low to calculate their $T_M$, shown in Fig. 4(d). We find all of them are metallic.

We further investigate the selected magnets, shown in Fig. 5. Except for $CrS_2$ 1T', all of them contain 3D-like networks that not only increase the number of the HEs but also eliminate the dimensional effect as their spin lattices are not strictly 2D. $J(r)$ of the ML forms and the bulk forms are similar. We find $Cr_2S_3$ 1T1T, $CrS_2$ 1T' and $Cr_3Te_4$ 1T2H structures have FM ground states, while $Cr_2X_3$ 1T2H (X=S, Se, Te) structures have AFM ground states except for $Cr_2Te_3$ bulk which is FM. For the 1T2H structures, an interesting fact is that $J_1$ corresponding to the vertical HE connecting the sublayers, is positive, which can cause an inter-sublayer AFM ground state. $J_2$ whose $\vec{r}_2$ is parallel to the plane and is negative, so inside the sublayer it is FM. Unlike the AA stack of the sublayers in the 1T2H structures, the 1T1T structure has sublayers AB stacked so FM is stabilized.

Then we study ML $CrS_2$. The 1T structure is AFM. Under the Peierls distortion, a new structure called 1T' is obtained with a bit higher formation energy, shown in Fig. 5(d). The 1T' structure is experimentally observed by chemical vapor deposition [38] and predicted to be a half-metal with $T_C$ ~ 1000K [39]. Rather than comparing the energy difference and using the mean-field approximation, we use *ASTDM*. Dimerization of Cr induces a symmetry breaking of the spin-lattice. $J_1$ splits into three HEs called $J_{11}, J_{12}, J_{13}$, sorted by the distance. $J_2$ splits into three HEs and $J_3$ splits into two. The most important EI is the intradimer $J_{11}$ which is ~ 60meV, resulting in $T_C$ approaching 620K. We find that the magnetism disappears in the bulk form of 1T' as we have searched the non-magnetic bulk structure.

Lastly, we find the 1T2H $Cr_3Te_4$, which has also been proposed previously [22,26]. As it has three sublayers, there are two kinds of Cr with slightly different spins, Cr in the middle (b-Cr) and Cr by the sides (a-Cr). Each ML is 1.24nm, with a 3D-like network providing abundant HEs, as shown in Fig. 5(e). We identify nine intralayer HEs and two interlayer HEs below 0.8nm. $J_1 > 0$, corresponding to the on-top inter-sublayer EI of the 1T2H structure. $J_3 < 0$, helping to stabilize the inter-sublayer FM. $J_{2a}$ and $J_{2b}$ are negative, stabilizing the intra-sublayer FM. $J_{2a} < J_{2b}$, $J_{5a} < J_{5b}$, indicating direct exchange is higher in a-Cr, as electrons are more localized. For bulk, the negative large $J_{z0}$ ensure the interlayer FM order. $T_C$ is calculated to be 590K for the ML and 535K for the bulk. The high Curie temperatures stem from a large number of HEs inside the ML. For the same kind of spin-lattice, increasing the ratio of magnetic ions indeed helps enhance $T_C$, which is also reported in iron-based compounds $Fe_xGeTe_2$ [40].

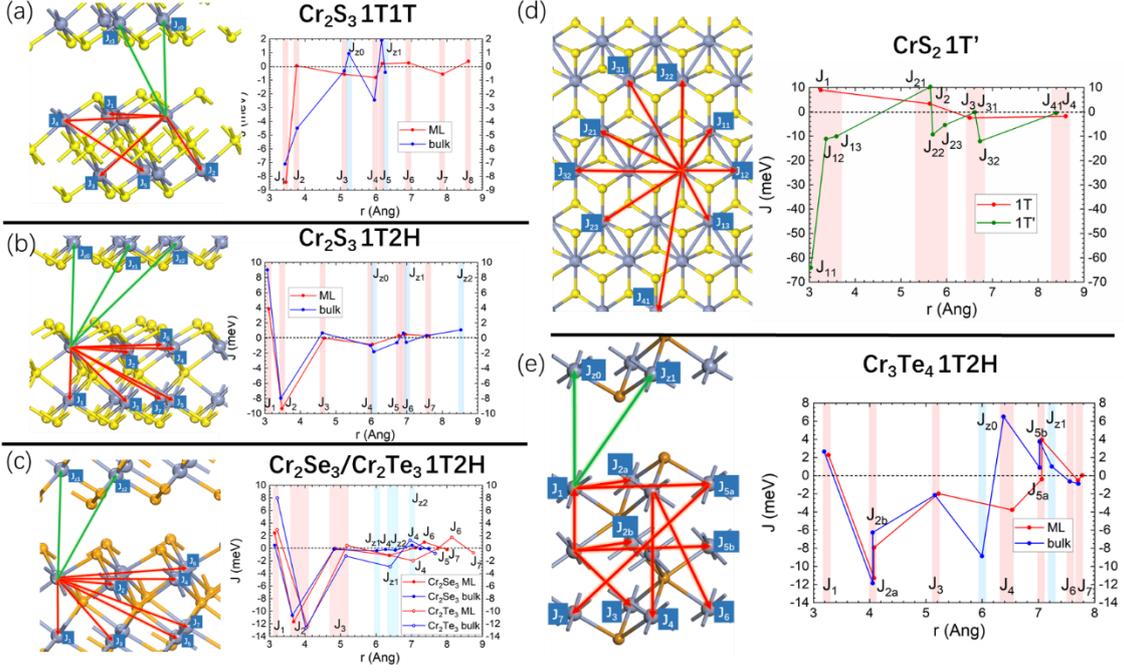

FIG. 5. Atomic structures and HEs for (a) $Cr_2S_3$ 1T1T, (b) $Cr_2S_3$ 1T2H, (c) $Cr_2Se_3$/ $Cr_2Te_3$ 1T2H, (d) ML $CrS_2$ 1T', (e) $Cr_3Te_4$ 1T2H. The intralayer HEs are marked in red, the interlayer HEs are marked in blue.

## D. Discussion

There are many interesting works to do with *ASTDM* in the future. First, a DFT+U test of various magnets can be done. The Hubbard U method is a widely used way to compensate for the underestimation of the Coulomb repulsion induced by electron correlation. For metals like $CrSe_2$ and $CrTe_2$, adding U reduces the overlap of the wavefunction and thereby reduce the overall exchanges, shown in Fig. S1. For semiconductors such as $CrCl_3$, $CrBr_3$, $CrI_3$, adding U increases the overall exchanges [29]. We are also aware that adding U can change MAE. For example, $CrGeTe_3$ is experimentally found 2D semiconductor with an out-of-plane anisotropy, while it is calculated to possess in-plane anisotropy (IMA) [6]. Adding U (0.2eV<U<1.7eV) to the 3d orbital of Cr changes from IMA to PMA, but adding U>1.7eV, it becomes interlayer AFM. $CrTe_2$ ML is calculated to be IMA, but adding U>4eV it is PMA [8].

For other exchanges than HEs, noncollinear magnetic configurations are usually needed for energy mapping method. How these configurations are automatically generated is an interesting topic in the future. If the spins are more than 1/2, higher order exchanges such as biquadratic exchanges should be taken into consideration. For example, biquadratic exchanges help stabilize the FM order in $NiCl_2$ monolayer [41]. If we only focus on transition temperature, short-range HEs are usually the largest terms to look for.

# IV. COCLUSION

The advantage of *ASTDM* is to straightforwardly investigate the Heisenberg exchange with complex spin-lattice and multiple Fermi surfaces. The energy mapping method has the advantage of efficient, one-off render of HEs for magnets with metastable states. The fine structure of HEs can be fundamental, especially considering the long-range HEs which enables us to observe the spin frustration in monolayer $CrSe_2$ and $CrTe_2$. Therefore, our work can be important for future magnetic structure investigations. *ASTDM* can also be interfaced with other high-throughput code to realize a quick check of large numbers of magnets, which can accelerate the pace of the magnet's data-mining and applications in the next-generation spintronic devices.

# ACKNOWLEDGEMENTS

H. L. thanks for the fruitful discussion with Prof. Stefan Blügel at Peter Grünberg Institute for magnetic exchange interaction calculations. This work is supported by the Beijing Natural Science Foundation (No. 2232055), the National Natural Science Foundation of China (No. 12204027), and the National Key R&D Program of China (No. 2022YFB3807200).

H. L. wrote *ASTDM* and conducted the main calculation, T. Y. performed random structure searches. H. L., J. R., and W. Z. supervised the project. H. L. wrote the paper and all authors contributed and commented on it.

# REFERENCES

1. I. Zutic, J. Fabian and S. Das Sarma, *Rev. Mod. Phys.* **76**, 323 (2004).
2. S. Wolf, D. Awschalom, R. Buhrman, J. Daughton, V. Von Molnar, M. Roukes, A. Chtchelkanova and D. Treger, *Science* **294**, 1488 (2001).
3. M. Gibertini, M. Koperski, A. Morpurgo and K. Novoselov, *Nat. Nanotech.* **14**, 408-419 (2019).
4. C. Gong and X. Zhang, *Science* **363**, 706 (2019).
5. B. Huang, G. Clark, E. Navarro-Moratalla, D. Klein, R. Cheng, K. Seyler, D. Zhong, E. Schmidgall, M. McGuire, D. Cobden, W. Yao, D. Xiao, P. Jarillo-Herrero and X. Xu, *Nature* **546**, 270 (2017).
6. C. Gong, L. Li, Z. Li, H. Ji, A. Stern, Y. Xia, T. Cao, W. Bao, C. Wang, Y. Wang, Z. Qiu, R. Cava, S. Louie, J. Xia and X. Zhang, *Nature* **546**, 265 (2017).
7. X. Zhang, Q. Lu, W. Liu, W. Niu, J. Sun, J. Cook, M. Vaninger, P. Miceli, D. Singh, S. Lian, T. Chang, X. He, J. Du, L. He, R. Zhang, G. Bian and Y. Xu, *Nat. Comm.* **12**, 2492 (2021).
8. L. Meng, Z. Zhang, M. Xu, S. Yang, K. Si, L. Liu, X. Wang, H. Jiang, B. Li, P. Qin, P. Zhang, J. Wang, Z. Liu, P. Tang, Y. Ye, W. Zhou, L. Bao, H. Gao and Y. Gong, *Nat. Comm.* **12**, 809 (2021).
9. D. J. O'Hara, T. Zhu, A. H. Trout, A. S. Ahmed, Y. Luo, C. Lee, M. R. Brenner, S. Rajan, J. A. Gupta, D. W. McComb and R. K. Kawakami, *Nano Lett.* **18**, 3125 (2018).

10. Y. Sun, Q. Tan, X. Liu, Y. Gao and J. Zhang, *J. Phys. Chem. Lett.* **10**, 3087-3093 (2019).

11. A. Szilva, Y. Kvashnin, E. A. Stepanov, L. Nordstrom, O. Eriksson, A. I. Lichtenstein and M. I. Katsnelson, *Rev. Mod. Phys.* **95**, 035004 (2023).

12. A. I. Liechtenstein, M. I. Katsnelson, V. P. Antropov and V. A. Gubanov, *J. Magn. Magn. Mater.* **67**, 65-74 (1987).

13. M. Pajda, J. Kudrnovsky, I. Turek, V. Drchal and P. Bruno, *Phys. Rev. B* **64**, 174402 (2001).

14. P. Bruno, *Phys. Rev. Lett.* **90**, 087205 (2003).

15. X. Wan, Q. Yin and S. Y. Savrasov, *Phys. Rev. Lett.* **97**, 266403 (2006).

16. H. Xiang, C. Lee, H. Koo, X. Gong and M. Whangbo, *Dalton Trans.* **42**, 823 (2013).

17. W. Heisenberg, *Z. Phys.* **49**, 619–636 (1928).

18. R. E. Prange and V. Korenman, *Phys. Rev. B* **19**, 4691 (1979).

19. B. Zimmermann, G. Bihlmayer, M. Böttcher, M. Bouhassoune, S. Lounis, J. Sinova, S. Heinze, S. Blügel and B. Dupé, *Phys. Rev. B* **99**, 214426 (2019).

20. H. Yang, A. Thiaville, S. Rohart, A. Fert and M. Chshiev, *Phys. Rev. Lett.* **115**, 267210 (2015).

21. N. Miao, B. Xu, L. Zhu, J. Zhou and Z. Sun, *J. Am. Chem. Soc.* **140**, 2417-2420 (2018).

22. Y. Zhu, X. Kong, T. Rhone and H. Guo, *Phys. Rev. Mater.* **2**, 081001 (2018).

23. D. Torelli, K. Thygesen and T. Olsen, *2D Mater.* **6**, 045018 (2019).

24. D. Torelli, H. Moustafa, K. W. Jacobsen and T. Olsen, *npj Comput. Mater.* **6**, 158 (2020).

25. S. Tiwari, J. Vanherck, M. Van de Put, W. Vandenberghe and B. Soree, *Phys. Rev. Research.* **3**, 043024 (2021).

26. A. Kabiraj, M. Kumar and S. Mahapatra, *npj Comput. Mater.* **6**, 35 (2020).

27. M. Horton, J. Montoya, M. Liu and K. Persson, *npj Comput. Mater.* **5**, 64 (2019).

28. F. Walsh, M. Asta and L. Wang, *npj Comput. Mater.* **8**, 186 (2022).

29. A. Kartsev, M. Augustin, R. Evans, K. Novoselov and E. Santos, *npj Comput. Mater.* **4**, 57 (2018).

30. M. T. Beal-Monod, *Phys. Rev. B* **36**, 8835 (1987).

31. Y. Zhu, Y. Pan, L, Ge, J. Fan, D. Shi, C. Ma, J. Hu and R. Wu, *Phys. Rev. B* **108**, L041401 (2023).

32. M. A. Ruderman and C. Kittel, *Phys. Rev.* **96**, 99 (1954).

33. H. Wang, H. Lu, Z. Guo, A. Li, P. Wu, J. Li, W. Xie, Z. Sun, P. Li, H. Damas, A. M. Friedel, S. Migot, J. Ghanbaja, L. Moreau, Y. Fagot-Revurat, S. Petit-Watelot, T. Hauet, J. Robertson, S. Mangin, W. Zhao and T. Nie, *Nat. Comm.* **14**, 2483 (2023).

34. B. Li, Z. Wan, C. Wang, P. Chen, B. Huang, X. Cheng, Q. Qian, J. Li, Z. Zhang, G. Sun, B. Zhao, H. Ma, R. Wu, Z. Wei, Y. Liu, L. Liao, Y. Yu, Y. Huang, X. Xu, X. Duan, W. Ji and X. Duan, *Nat. Mater.* **20**, 818-825 (2021).

35. C.J. Pickard and R. Needs, *J. Phys.: Condens. Matter.* **23**, 053201 (2011).

36. M. Segall, C. Pickard, P. Hasnip, M. Probert, K. Refson and M. Payne, *Z. Kristallogr. Cryst. Mater.* **220**, 567-570 (2005).

37. P. Larsen, M. Pandey, M. Strange and K. Jacobson, *Phys. Rev. Mater.* **3**, 034003 (2019).

38. M. Habib, S. Wang, W. Wang, H. Xiao, S. Obaidulla, A. Gayen, Y. Khan, H. Chen and M. Xu, *Nanoscale* **11**, 20123-20132 (2019).
39. K. Chen, J. Deng, Y. Yan, Q. Shi, T. Chang, X. Ding, J. Sun, S. Yang and J. Z. Liu, *npj Comput. Mater.* **7**, 79 (2021).
40. J. Seo, D. Kim, E. An, K. Kim, G. Kim, S. Hwang, D. Kim, B. Jang, H. Kim, G. Eom, S. Seo, R. Stania, M. Muntwiler, J. Lee, K. Watanabe, T. Taniguchi, Y. Jo, J. Lee, B. Min, M. Jo, H. Yeom, S. Choi, J. Shim and J. Kim, *Sci. Adv.* **6**, eaay8912 (2020).
41. J. Ni, X. Li, D. Amoroso, X. He, J. Feng, E. Kan, S. Picozzi and H. Xiang, *Phys. Rev. Lett.* **127**, 247204 (2021).

**Supplementary Materials: Automatic Calculation of the Transition Temperatures for two-dimensional Heisenberg type Magnets**

Haichang Lu[1,2,3]*, Tai Yang[1], Zhimei Sun[4], John Robertson[3]* and Weisheng Zhao[1,2]*

[1]*Fert Beijing Institute, MIIT Key Laboratory of Spintronics, School of Integrated Circuit Science and Engineering, Beihang University, Beijing, 100191, China2*

[2]*National Key Lab of Spintronics, Institute of International Innovation, Beihang University, Yuhang District, Hanzhou, 311115, China*

[3]*Engineering Department, Cambridge University, Cambridge CB2 1PZ, UK*

[4]*School of Materials Science and Engineering, Beihang University, Beijing 100191, China*

**This PDF file includes:**

Sections I to IV

Figs. S1 to S22

Table S1 and S2

References

# I First-principle calculations

## 1. Calculation setting

The cutoff energy is set to 570eV. The energy tolerance is $10^{-6}$ eV/ion. $k_c$ varies from 60 to 130 (see the definition of $k_c$ in **SI III**). For itinerant magnets, $k_c$ is higher. There is no k-point sampling along the vacuum direction. The generalized gradient approximation (GGA) form of Perdew-Burke-Ernzenhof (PBE) [1] is employed as the exchange-correlation functional. Tkatchenko-Scheffler (TS) correction [2] is applied for layered materials to interpret van der Waals interaction. No Hubbard U correction is added unless stated. For systems with magnetic anisotropy or spin-orbital coupling effect is considered, noncollinear spin calculations are performed. *ASTDM* can integrate with the Cambridge Serial Total Energy Package (CASTEP) [3] and the Vienna ab initio simulation package (VASP) [4], where we use the latter in this work.

## 2. k-point test

Fig. S1 shows the k-point test of metallic $CrSe_2$ and $CrTe_2$ as the showcase. The k-point density $k_c$ is defined as the lattice constant $l$ multiplied by the k-point density $k$ in each direction, so $k = floor(k_c/l)$. We find that metals usually require larger $k_c$, especially in two dimensions For instance, adopt $k_c \sim 130$ for ML and $k_c \sim 80$ for bulk $CrTe_2$. Whereas $k_c \sim 50$ is enough for most insulating magnets.

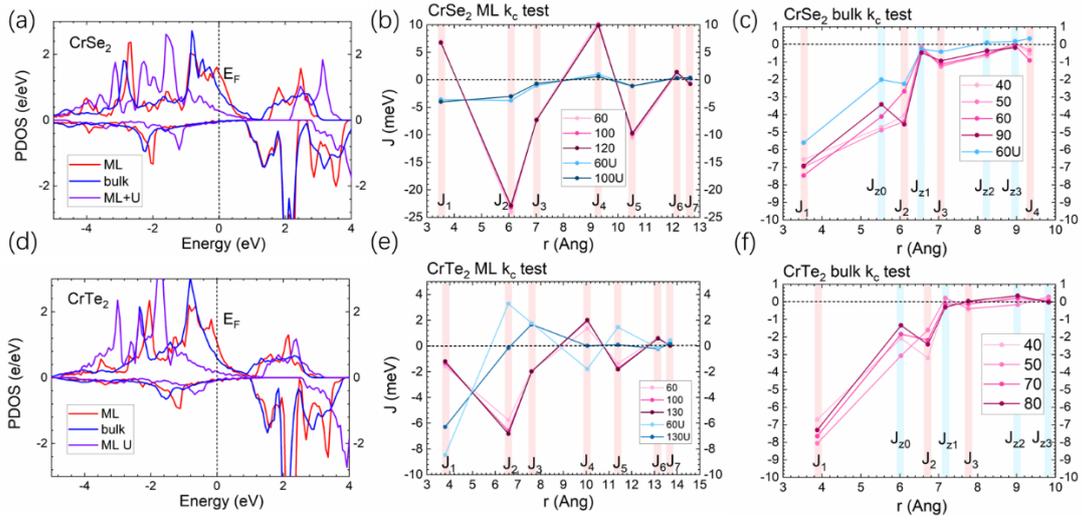

FIG. S1. Partial density of states (PDOS) of Cr bands in (a) $CrSe_2$, (b) $CrTe_2$ monolayer (ML) marked in blue, bulk in red, ML with Hubbard U correction (U=4.6eV, J=0.6eV, same as in ref [5]) in violet. The k-point tests of (c) ML, (d) bulk $CrSe_2$, and (e) ML, (f) bulk $CrTe_2$. Intralayer Heisenberg exchanges are marked in light pink and interlayer Heisenberg exchanges are marked in light blue.

# II Monte Carlo calculations

## 1. Monte Carlo sampling

The Monte Carlo sampling can be done either classically or semi-classically, under the framework of the Heisenberg model. First, a random site $i$ is selected to generate its trial direction.

For classical sampling, we adopt sphere point picking to generate a random direction. The location of a point on a unit sphere can be written as the spherical coordinate $(\theta, \varphi)$, where $\theta \in [0, \pi]$, $\varphi \in [0, 2\pi)$. The area element is

$$d\Omega = sin\theta d\theta d\varphi$$

The random point picking on the sphere surface should be in the manner that each area element is picked with equal probability. Therefore, the spherical surface can be mapped into a unit square UV so that the random sphere surface point choice is equivalent to the random point choice on the square.

$$\theta = \arccos(1 - 2v)$$

$$\varphi = 2\pi u$$

Then $u \in [0,1)$, $v \in [0,1]$ can be randomly selected.

For semiclassical sampling, we adopt local quantization of the chosen spin on site $i$. The Heisenberg Hamiltonian for site $i$ is equivalent to a single-spin quantum Hamiltonian,

$$H_{mag} = E_0 + E_{j \neq i} + E_{mae} + \vec{s}_i \cdot \sum_j J_{ij} \vec{s}_j$$

where the rest spins interacting with the i$^{th}$ spin can be treated as an effective magnetic field $\vec{B}_{eff} = -\sum_j J_{ij} \vec{s}_j$, so that $\vec{s}_i$ is quantized along $\vec{B}_{eff}$. Very rarely if $\vec{B}_{eff} = 0$, the new $\vec{s}_i$ is oriented randomly. Otherwise, the quantum number $m$ depicting the projected spin is randomly selected from $-s, -s+1, \ldots, s-1, s$, then the local spherical coordinate can be generated by

$$\theta' = \arccos\left(\frac{m}{s}\right)$$

$$\varphi' = 2\pi u$$

where $u \in [0,1)$ is randomly selected. Then the new spherical coordinate $(\theta, \varphi)$ of $\vec{s}_i$ can be deducted from

$$\frac{\vec{s}_i}{s_i} = \begin{pmatrix} sin\varphi_b & cos\theta_b cos\varphi_b & sin\theta_b cos\varphi_b \\ -cos\varphi_b & cos\theta_b sin\varphi_b & sin\theta_b sin\varphi_b \\ 0 & -sin\theta_b & cos\theta_b \end{pmatrix} \begin{pmatrix} sin\theta' cos\varphi' \\ sin\theta' sin\varphi' \\ cos\theta' \end{pmatrix} = \begin{pmatrix} sin\theta cos\varphi \\ sin\theta sin\varphi \\ cos\theta \end{pmatrix}$$

where $(\theta_b, \varphi_b)$ is the direction of $\vec{B}_{eff}$.

The initial spin setting for each Monte Carlo calculation is randomly oriented (paramagnetic) to avoid spin freeze at low temperatures in the case of the semiclassical treatment for the spin frustrated system. Though the semiclassical Monte Carlo (SMC) neglects the low-temperature entanglement and is unable to conduct cluster quantum flipping, it largely improved from classical Monte Carlo (CMC) as it sample spin continuously, which should only be in the case of $S \to \infty$. When using SMC, a scaling is applied to the exchange interactions to be consistent with CMC:

$$S(S+1)J^{SMC} = S^2 J^{CMC}$$

**2. Heat capacity, Magnetic susceptibility and Normalized Entropy**

Observing the heat capacity or the magnetic susceptibility is a good way to determine the transition temperatures, especially for antiferromagnets without explicit magnetic moments. The heat capacity is defined as

$$C_V^1 = \frac{\partial E}{\partial T}$$

where $E$ is the magnetic energy. The heat capacity can also be calculated using the fluctuation-dissipation theorem

$$C_V^2 = \frac{\overline{E^2} - \overline{E}^2}{k_B T^2}$$

where $k_B$ is the Boltzmann constant. These two methods must give equivalent results in the thermodynamic limit if the size of the lattice and the loop number are close to infinity. However, the fluctuation-dissipation theorem is more accurate for a finite-size lattice and a limited number of loops. Therefore, we also calculate the zero-field magnetic susceptibility as

$$\chi = \frac{\overline{M^2} - \overline{M}^2}{k_B T}$$

where $M$ is the magnetic moment.

To obtain the normalized entropy, we collect the direction of each spin in the lattice for each Monte Carlo loop. The direction can then be mapped into a point in the UV. We divided the unit square UV into a $G \times G$ grid, so $du = dv = \frac{1}{G}$. For a lattice with $N$ spin sites, the probability function of the i$^{th}$ site $p_i(u,v)$ is

$$p_i(u,v) = \frac{number\ of\ points\ inside\ dudv\ at\ (u,v)\ for\ site\ i}{N_l}$$

where $N_l$ is the number of loop. Finally, we normalized the entropy, defined as:

$$S = -k_B \frac{\sum_i \sum_{u,v} p_i(u,v) ln p_i(u,v)}{\ln(G^2) \times N}$$

$S = 1$ corresponds to the disordered paramagnetic state, $S = 0$ corresponds to the ordered magnetic state. We can visualize the most probable direction of any site by the output of $(u, v)$ which has the max $p_i(u,v)$. Fig. S2 shows CrI$_3$ ML as an example to illustrate the transition. Below T$_C$=70K, the spin vectors are parallel to the z direction, $S < 1$. While upon the transition, the spins orient randomly, $S \approx 1$, the magnetization drops to zero.

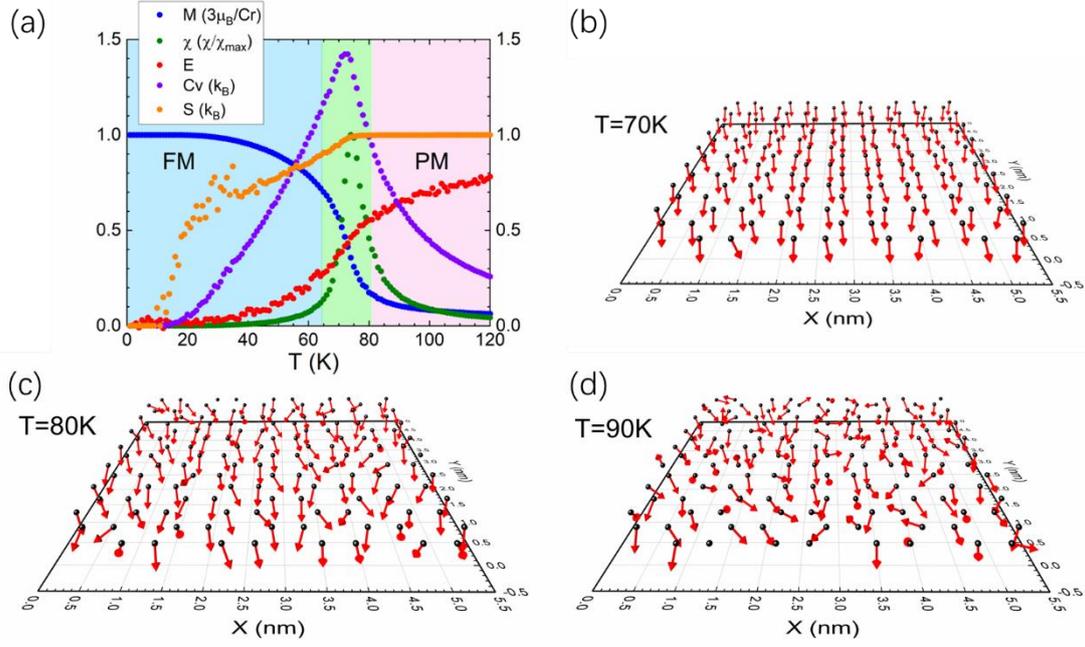

FIG. S2. Monte Carlo calculation of CrI$_3$ monolayer. (a) Temperature-dependent normalized magnetization, magnetic susceptibility, energy (FM=0, PM=1), heat capacity, and entropy. Loop range is $10^5 \sim 5 \times 10^5$. The lattice size is $20 \times 20$. The FM region is marked in light blue and the PM region is marked in light pink, separated by phase transition region marked in green. Visualization of the spin vectors during the phase transition, at (b) 70K, (c) 80K, and (d) 90K. The black dots are Cr, and the red arrows are the spin vectors, pointing toward the most probable direction.

### 3. Finite-size test

We use the ML CrI$_3$ as the showcase. The magnetic moment and the heat capacity are calculated with different sizes of the lattice. The total loop is 50000 while we discard the first 10000 loops to ensure the system reaches the equilibrium. The insufficient size of the lattice causes an inaccurate interpretation of the paramagnetic state as well as an overestimation of the transition temperature. The magnetization test in Fig. S2(a) shows from $30 \times 30$, the errors are acceptable. The semiclassical treatment shows freeze of the spins due to quantum restriction at low temperatures. Additionally, the heat capacity of the classical MC near 0K is $k_B$, but it should be zero by the third law of thermodynamics. The semiclassical treatment solves this problem. Heat capacity calculated by the fluctuation-dissipation theorem ($C_V^2$) is better than the energy gradient $C_V^1$. We also note that larger lattice requires more loops to find the equilibrium state. A loop range of 10000~50000 is not enough for a lattice size of more than 20. Therefore, a loop test is necessary.

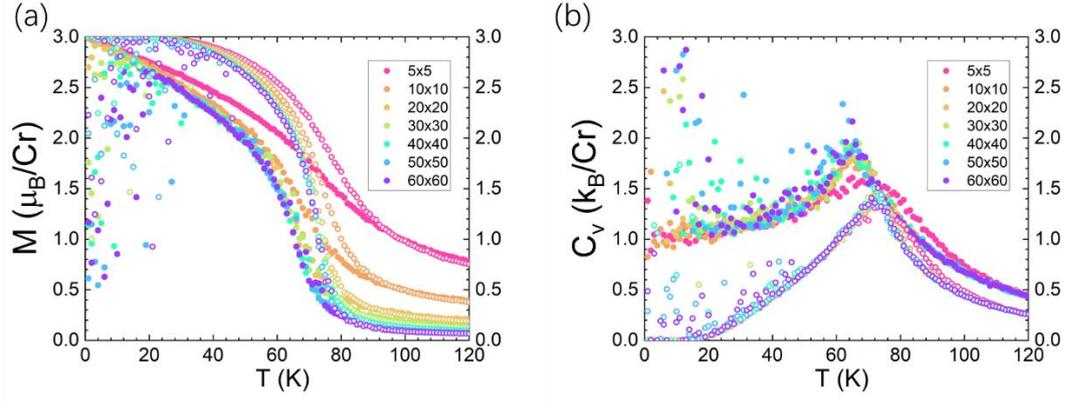

FIG. S3. Finite-size test of the Monte Carlo calculation in CrI$_3$ monolayer, from $5 \times 5$ to $100 \times 100$. The loop range is 10000~50000. The solid dots stand for classical treatments while the hollow dots stand for semiclassical treatments. We test (a) magnetization and (b) heat capacity using the fluctuation-dissipation theorem.

### 4. Loop test

We still use the ML CrI$_3$ as the showcase. The magnetic moment, magnetic susceptibility, and heat capacity are calculated with different loop numbers. The lattice size is $30 \times 30$. We find that a loop range of $10^5$~$5 \times 10^5$ is enough.

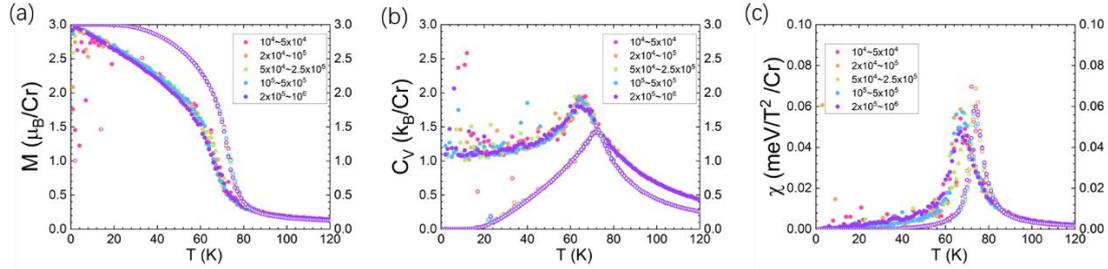

FIG. S4. Loop test of the Monte Carlo calculation in CrI$_3$ monolayer, from $10^4$~$5 \times 10^4$ to $2 \times 10^5$~$10^6$. The lattice size is $30 \times 30$. The solid dots stand for classical treatments while the hollow dots stand for semiclassical treatments. We test the (a) magnetization, (b) heat capacity using the fluctuation-dissipation theorem, and (c) magnetic susceptibility.

### 5. 2D spin-lattice with in-plane anisotropy

We use the ML CrGeTe$_3$ as the showcase. Though the experiment reveals it has out-of-plane anisotropy, the DFT without U correction shows in-plane anisotropy. The lattice size is $30 \times 30$. Loop range is $10^5$~$5 \times 10^5$. Without the external magnetic field, there is no long-range order. Due to the finite-size effect, we can have the wrong magnetization. Adding the field can solve the problem. We choose 0.1T and 1T fields, both out-of-plane and in-plane. The magnetic field energy of 1T field is ~0.058meV per magnetic moment, which can sustain an easy axis anisotropy. The effect of the field on the transition temperature is small compared to exchange interactions, the magnetization curve drops at a bit higher temperature with the field.

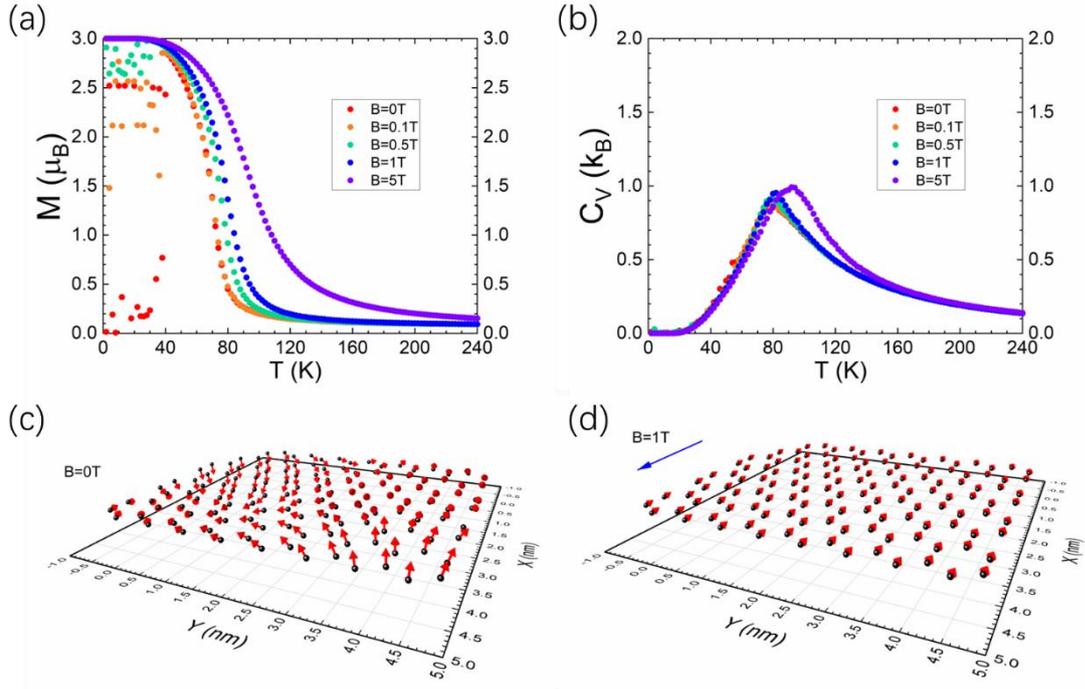

FIG. S5. Monte Carlo calculation of CrGeTe$_3$ monolayer under in-plane magnetic field: (a) magnetization, (b) heat capacity. Loop range is $10^5 \sim 5 \times 10^5$. The lattice size is $30 \times 30$. We show the spin configuration at 6K of (c) zero-field, (d) 1T magnetic field.

### 6. Monte Carlo setting

We set the size of the lattice to vary with the type of spin-lattice. Usually, the total number of spin sites is ~1800. For example, we set $30 \times 30 \times 1$ for 2D CrI$_3$ and $12 \times 12 \times 12$ for 3D CrI$_3$, $40 \times 40 \times 1$ for 2D CrTe$_2$ and $15 \times 15 \times 8$ for 3D CrTe$_2$. The loop range is from $10^5$ to $5 \times 10^5$ to remove the errors from the initial setting and capture the thermal equilibrium state. For 2D spin-lattice with in-plane anisotropy, we add an in-plane field of 1T as an effective easy axis anisotropy to ensure the long-range order exists.

### III. Guides of *ASTDM*

**The input structure file** is written in a text file containing the information on the lattice vectors and the ion coordinate. Below is the relaxed structure of the CrI$_3$ monolayer as an example:

```
#Lattice_Constant#
    6.081632099972371    -3.511231930031315     0.000000000000000
    0.000000000000000     7.022463860062633     0.000000000000000
    0.000000000000000     0.000000000000000    29.999999889085267
#Lattice_Constant_end#
```

```
#Ion_frac_coordinate#
 Cr  0.3333333333333334   0.6666666666666667   0.4999999999999999   3.0000000000
 Cr  -0.3333333333333334  -0.6666666666666667  -0.4999999999999999  3.0000000000
 I   0.3556232621965423   -0.0000000000000001  0.4478635151832834   0
 I   0.0000000000000002   0.3556232621965425   0.4478635151832834   0
 I   -0.3556232621965425  -0.3556232621965424  0.4478635151832834   0
 I   0.0000000000000002   -0.3556232621965423  -0.4478635151832834  0
 I   -0.3556232621965425  -0.0000000000000002  -0.4478635151832834  0
 I   0.3556232621965423   0.3556232621965425   -0.4478635151832834  0
#Ion_frac_coordinate_end#
```

The first block is the lattice constant, where three lattice vectors should be keyed in successively. The second block contains ion information. The first row is the ion species. Then the second, third, and fourth rows are the fractional coordinates. The last row is the size of the magnetic moment, in the unit of $\mu_B$.

**The parameter file** contains all the necessary coefficients to set up the calculation for the current version. Below are the keywords:

*input*: the name of the input structure file.

*method*: DFT code to calculate total energy, 0 for CASTEP, 1 for VASP.

*non_mag_spin_max*: the maximum spin moment of an ion to be treated as a non-magnetic ion $\Delta s$, the unit is $\mu_B$.

*Ex_In_max_r*: the cutoff spatial distance $r_c$ of the exchange interactions, in the unit of Å. Only $J_{ij}$ satisfying $r_{ij} = |\vec{r}_i - \vec{r}_j| < d_c$ are taken into counted.

*same_spin_tor*: the tolerance of the spin moments $\delta s$ to be treated as the same, in the unit of $\mu_B$.

*same_r_tor*: the tolerance of the distances $\delta r$ to be treated as the same, in the unit of Å.

*search_scale*: number of times the primitive lattice needs to expand in the three directions to encapsulate all exchange interactions counted. For a given spin site, high-order neighbors can be located far away so expanding the cell is required. The choice of it should correspond to $d_c$.

*vdW*: whether van der Waals corrections are adopted.

*XC*: the exchange-correlation functional used, can be either generalized gradient approximation (GGA) or Local Density Approximation (LDA).

*SOC*: whether the spin-orbit coupling effect is considered.

*vacuum*: the direction of the vacuum, if there is any. It can be x, y, z for two-dimensional materials, xy, xz, yz for one-dimensional materials (chain), and xyz for zero-dimensional materials (quantum dots).

*mae*: whether magnetic anisotropic energy is considered.

*mae_axis*: the direction of the easy axis.

*k_point_criteria*: $k_c$ is defined as the lattice constant $l$ multiplied by the k-point density $k$ in each direction. Then $k = floor(k_c/l)$.

*io_spin_tor*: tolerance between the initial spin moment and the DFT relaxed spin moment of each magnetic ion. This is to examine whether the chosen meta-stable state exists.

*mc_type*: 0 for classical Monte Carlo, 1 for semiclassical Monte Carlo.

*iter_type*: 0 for lattice iteration, 1 for random site sampling.

*mc_lattice*: the size of the spin-lattice for Monte Carlo calculation.

*delta_T*: temperature step.

*T_no*: number of temperatures needed for Monte Carlo calculation.

*T_start*: the initial temperature.

*loop_range*: to capture the thermal equilibrium state in Monte Carlo calculation, the first few loops should be discarded. It gives the range to calculate the mean magnetic moments and entropies.

*grid*: the density of the two-dimensional $(u, v)$ grid to calculate the normalized entropy. A denser grid is more accurate.

*B*: the magnitude of the external magnetic field, the extra term $-\mu_s \vec{B} \cdot \sum_i \vec{s}_i$ is added to the Hamiltonian.

*B_th*: the Euler angle $\theta$ of the magnetic field.

*B_ph*: the Euler angle $\varphi$ of the magnetic field.

*mc_thread*: number of threads for parallel Monte Carlo calculation.

**Using the code:** *ASTDM* is written in C++ with OpenMP API and is very handy to use. Executing *ASTDM* in a directory with the input file and "parameter.txt" file will generate a subfolder called "Ex_In" containing folders of all necessary magnetic configurations in which there are DFT input files and run scripts. Users can revise the script to match the computer resources of their machines. There is another subfolder called "MAE" if magnetic anisotropy energy is considered, containing two folders where the initial moments are set to be parallel and perpendicular to the anisotropy axis. When all total energy calculations are finished, executing *ASTDM* again will trigger the collection of the total energies from the result files (*.castep file for CASTEP and OUTCAR file for VASP). The code will check whether the initial magnetic configurations are consistent with the output configurations. If passing the check, the exchange interactions and MAE will be deducted. Then it moves on to Monte Carlo calculation. The result of the calculations including spin-lattice analysis, exchange interactions analysis, configuration setup, DFT input generation, matrix construction, energy collection, exchange interactions/MAE extraction, and temperature-dependent magnetic moment and entropy will be printed out in 'ASTDM.log' file.

# IV. Material investigation

## 1. 2D magnets

TABLE S1: Comparison of experimental and calculated transition temperatures (Kelvin), MAE ($\mu$eV/magnetic ion) for 2D magnets, * for AFM magnets, # for metals, (sub) means adsorbed onto the substrate, (U) means adopting Hubbard U correction (U=4.6eV, J=0.6eV).

| Materials | ML/Bulk | Experiments | This work | MAE |
|---|---|---|---|---|
| $CrCl_3$ | ML | 17 [6] | 38 | -5.32 |
|  | Bulk | 27 [6] | 40 | -4.70 |
| $CrBr_3$ | ML | 34 [7] | 52 | -24.03 |
|  | Bulk | 47 [8] | 63 | -22.01 |
| $CrI_3$ | ML | 45 [9] | 71 | -24.80 |
|  | Bulk | 61 [9] | 73 | -34.37 |
| $CrSiTe_3$ | ML | 17 [10] | 59 | -6.43 |
|  | Bulk | 33 [10] | 78 | -33.41 |
| $CrGeTe_3$ | ML | 30 [11] | 80 | 62.25 |
|  | Bulk | 68 [11] | 134 | 61.43 |
| $MnPS_3$* | ML | 70 [12] | 192 | 5.74 |
|  | Bulk | 77 [12] | 205 | 5.68 |
| $MnPSe_3$* | ML | 74 [13] | 168 | 67.88 |
|  | Bulk | 74 [13] | 196 | 112.97 |
| $CrSe_2$# | ML | 60(sub) [5] | 290/250(U) | 96.76/161.46(U) |
|  | Bulk | 110 [5] | 495/300(U) | 44.56/14.03(U) |
| $CrTe_2$# | ML | 150(sub) [14] | 255/125(U) | 90.99/96.37(U) |
|  | Bulk | 300 [14] | 395 | 159.74 |

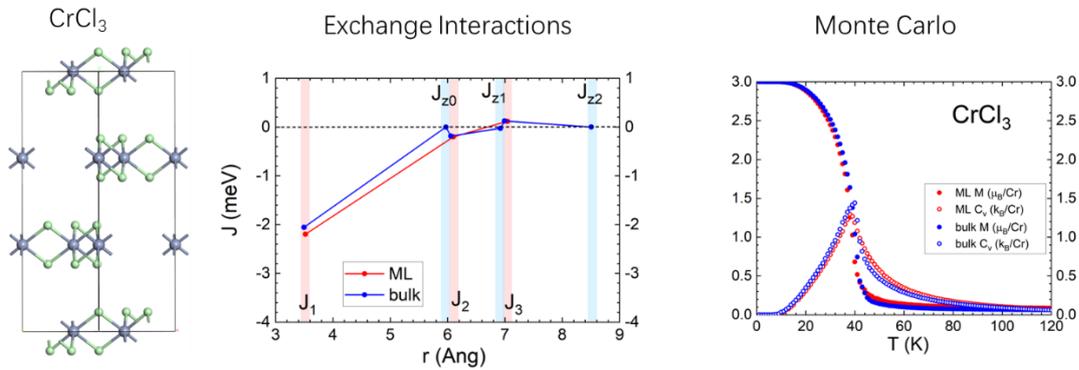

FIG. S6. Crystal structure, Heisenberg exchanges, temperature-dependent magnetic moment, and heat capacity of CrCl$_3$.

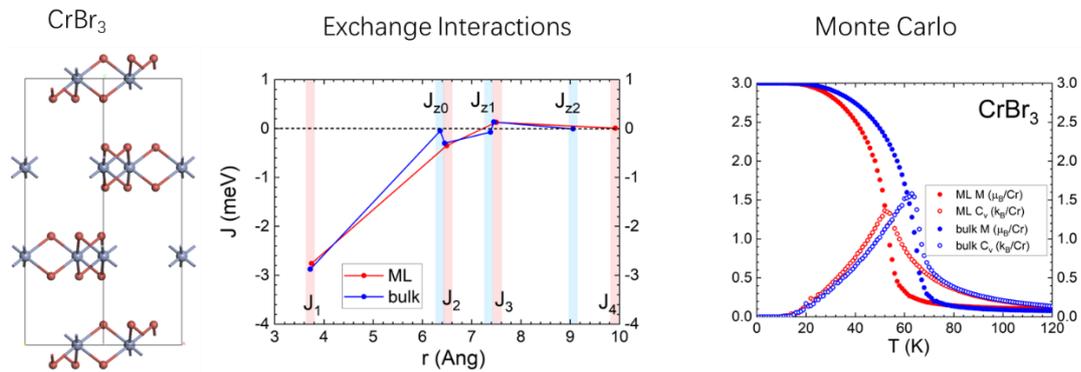

FIG. S7. Crystal structure, Heisenberg exchanges, temperature-dependent magnetic moment, and heat capacity of CrBr$_3$.

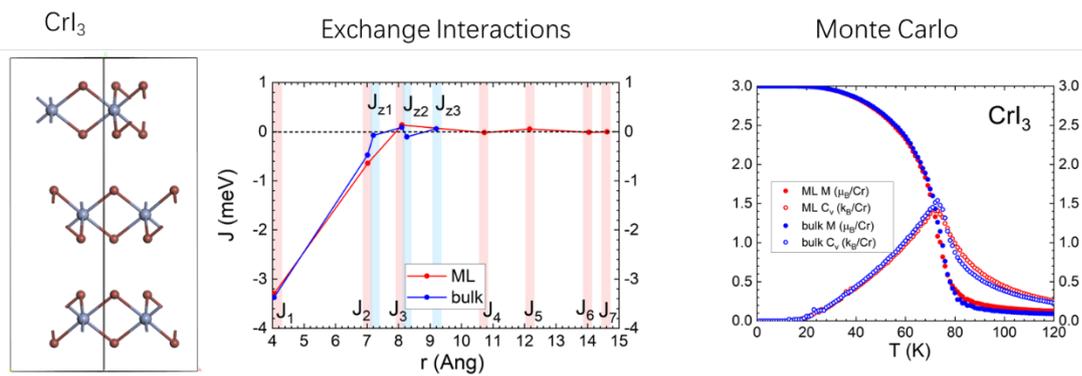

FIG. S8. Crystal structure, Heisenberg exchanges, temperature-dependent magnetic moment, and heat capacity of CrI$_3$.

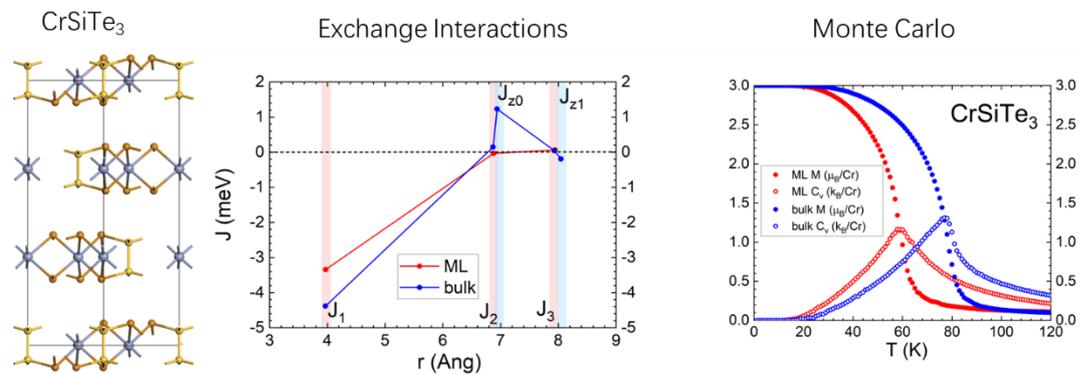

FIG. S9. Crystal structure, Heisenberg exchanges, temperature-dependent magnetic moment, and heat capacity of CrSiTe$_3$.

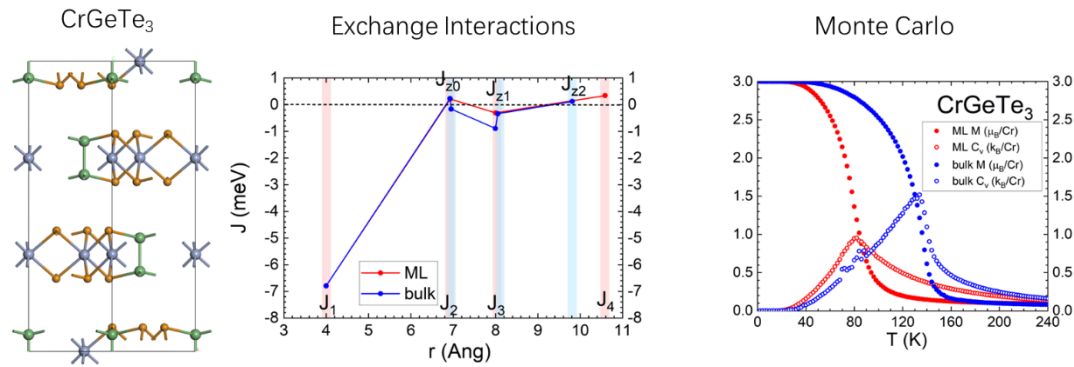

FIG. S10. Crystal structure, Heisenberg exchanges, temperature-dependent magnetic moment, and heat capacity of CrGeTe$_3$.

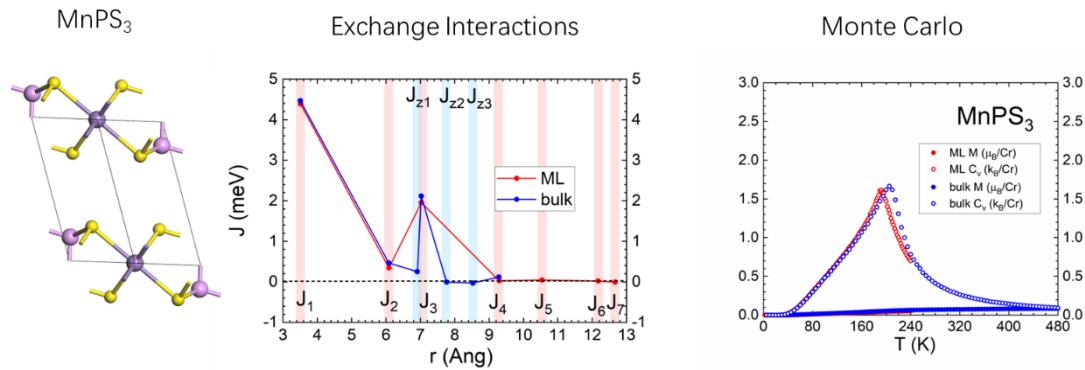

FIG. S11. Crystal structure, Heisenberg exchanges, temperature-dependent magnetic moment, and heat capacity of MnPS$_3$.

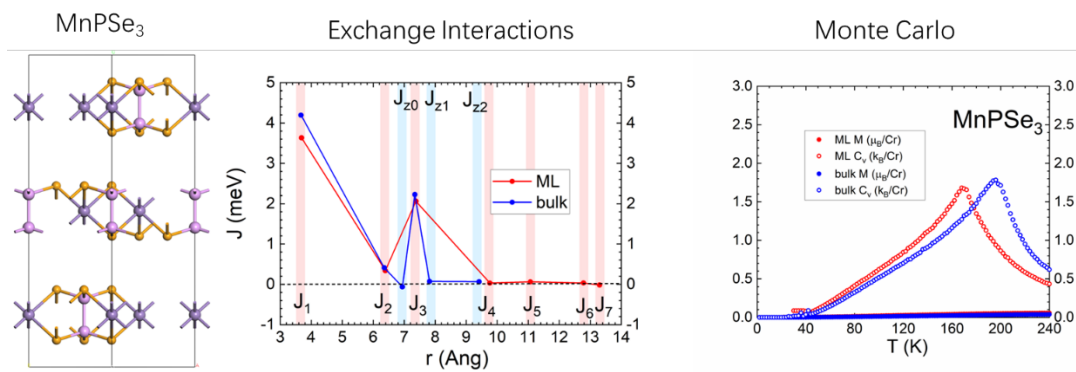

FIG. S12. Crystal structure, Heisenberg exchanges, temperature-dependent magnetic moment, and heat capacity of MnPSe$_3$.

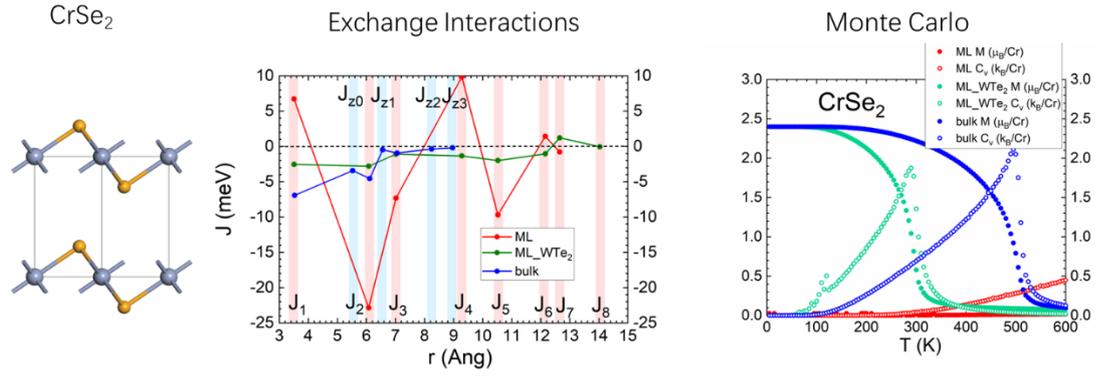

FIG. S13. Crystal structure, Heisenberg exchanges, temperature-dependent magnetic moment, and heat capacity of $CrSe_2$.

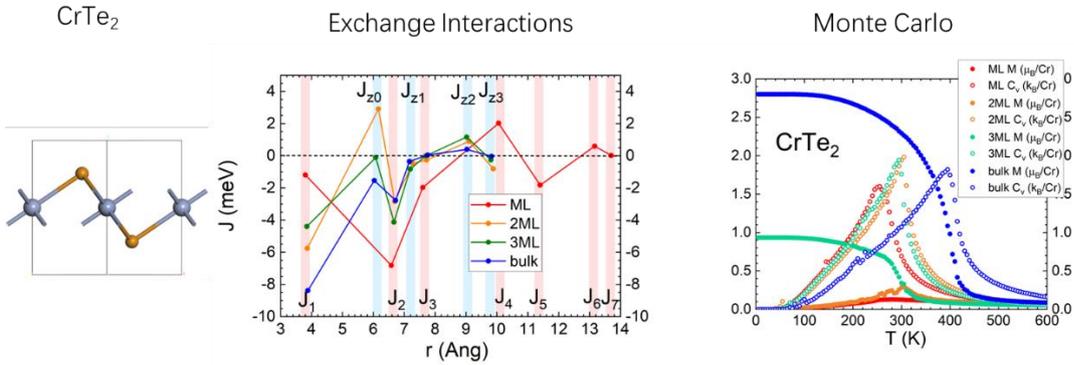

FIG. S14. Crystal structure, Heisenberg exchanges, temperature-dependent magnetic moment, and heat capacity of $CrSe_2$.

## 2. 3D magnets

TABLE S2: Comparison of experimental and calculated transition temperatures (Kelvin), MAE ($\mu$eV/magnetic ion) for 3D magnets, * for AFM magnets, # for metals.

| Materials | Experiments | This work | MAE |
|---|---|---|---|
| MnO*# | 120 [15,16] | 170 | 0 |
| MnS*# | 133 [17,18] | 155 | 0 |
| MnSe*# | 123 [18] | 185 | -1.24 |
| MnTe*# | 283 [18] | 285 | 104.47 |
| MnBi# | 635 [19] | 850 | -0.89 |
| $CrO_2$# | 391 [20] | 680 | -7.91 |
| Cr$_2$O$_3$*# | 303 [18,21] | 325 | 1.52 |

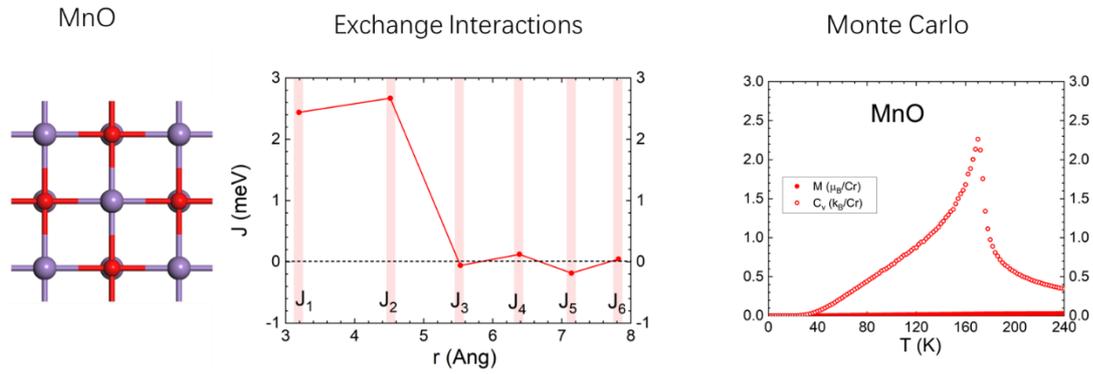

FIG. S15. Crystal structure, Heisenberg exchanges, temperature-dependent magnetic moment, and heat capacity of MnO.

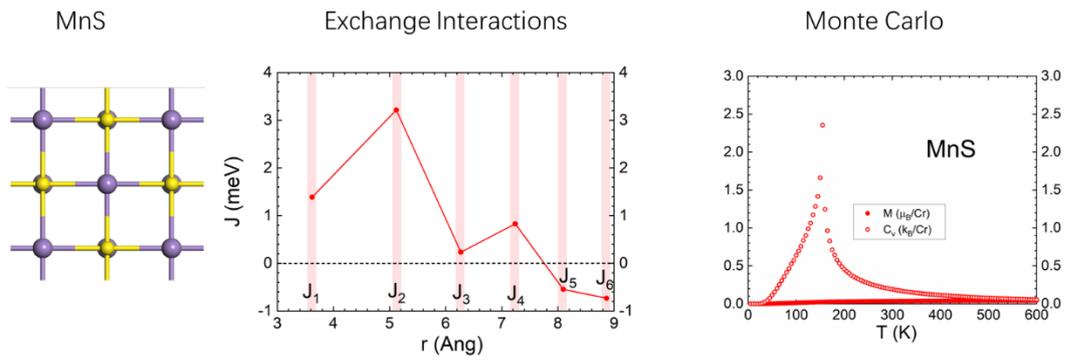

FIG. S16. Crystal structure, Heisenberg exchanges, temperature-dependent magnetic moment, and heat capacity of MnS.

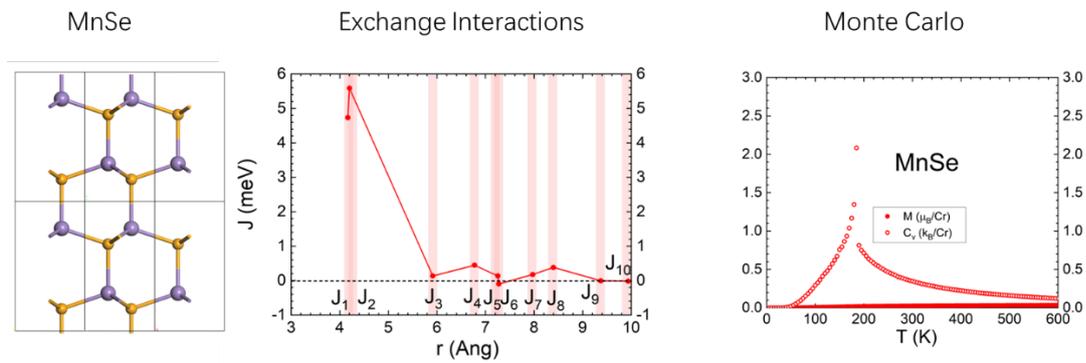

FIG. S17. Crystal structure, Heisenberg exchanges, temperature-dependent magnetic moment, and heat capacity of MnSe.

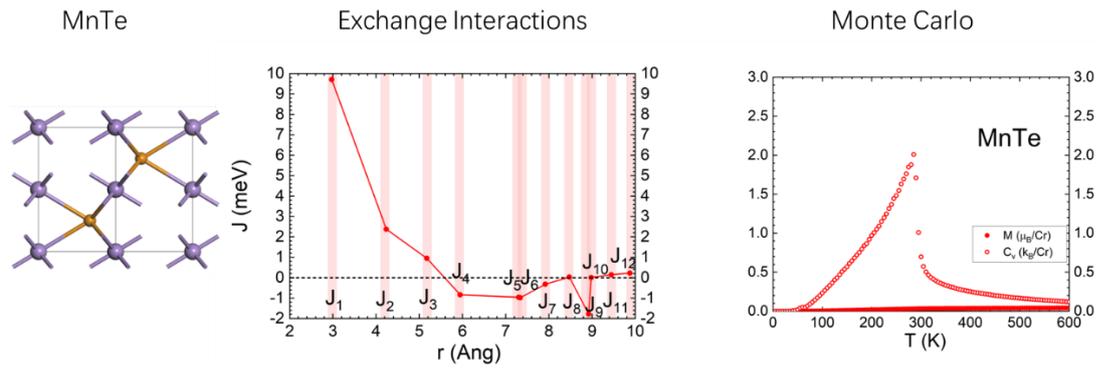

FIG. S18. Crystal structure, Heisenberg exchanges, temperature-dependent magnetic moment, and heat capacity of MnTe.

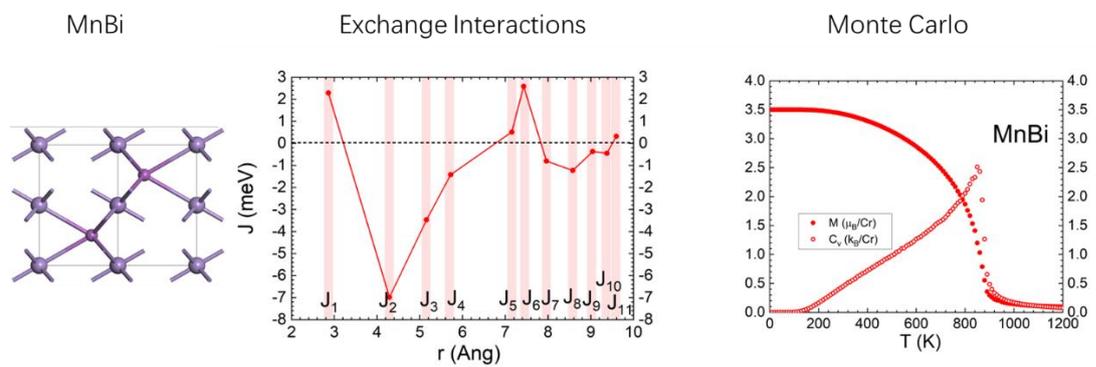

FIG. S19. Crystal structure, Heisenberg exchanges, temperature-dependent magnetic moment, and heat capacity of MnBi.

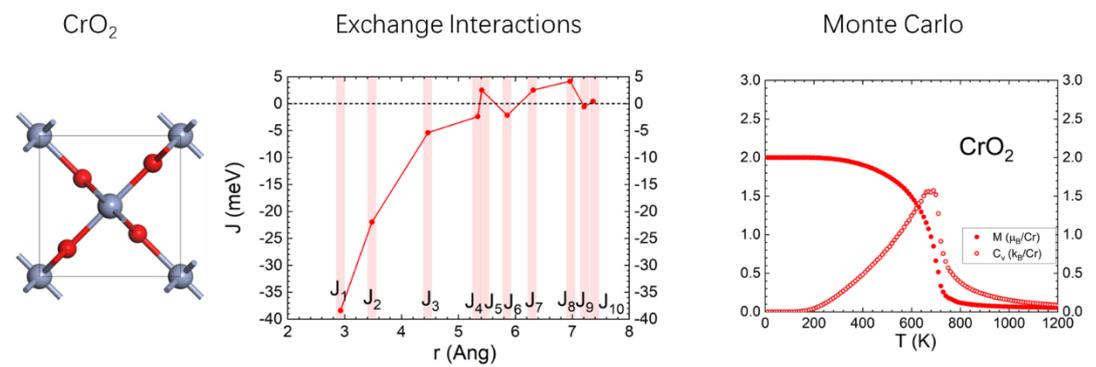

FIG. S20. Crystal structure, Heisenberg exchanges, temperature-dependent magnetic moment, and heat capacity of $CrO_2$.

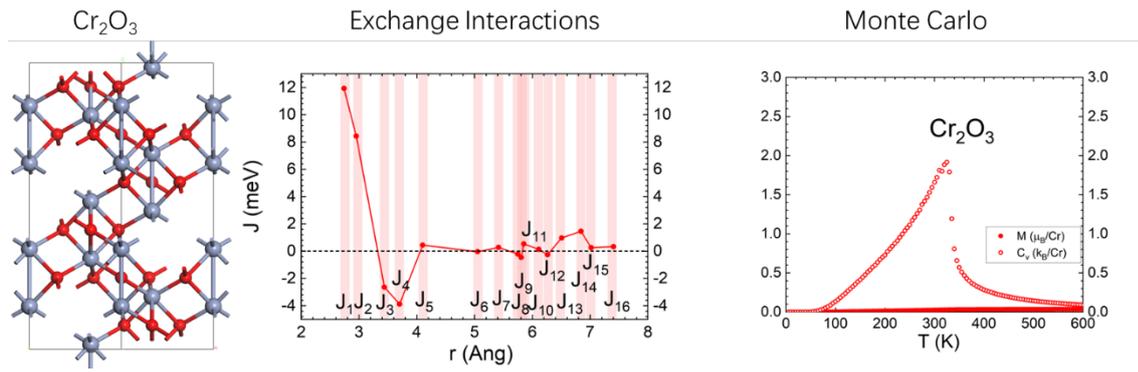

FIG. S21. Crystal structure, Heisenberg exchanges, temperature-dependent magnetic moment, and heat capacity of $Cr_2O_3$.

# V. Heterostructures

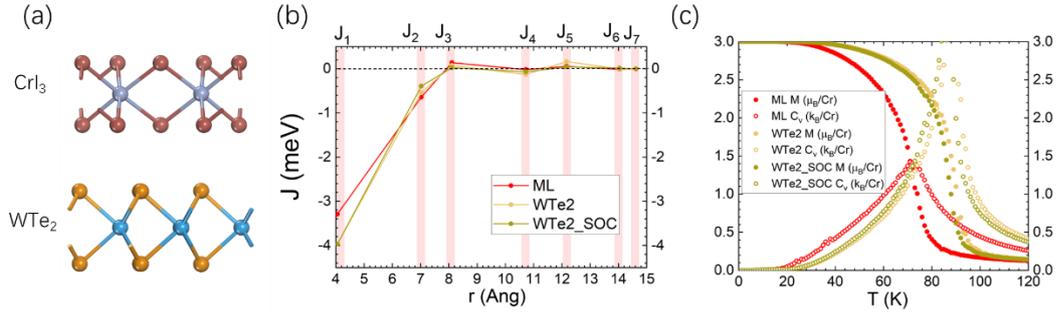

FIG. S22. (a) Atomic structure of the CrI$_3$-WTe$_2$ heterostructure, (b) HEs, and (c) M, C$_V$ of suspended ML CrI$_3$, the heterostructure.

**References**

1. J. Perdew, K. Burke and M. Ernzerhof, *Phys. Rev. Lett.* **77**, 3865 (1996).
2. A. Tkatchenko and M. Scheffler, *Phys. Rev. Lett.* **102**, 073005 (2009).
3. M. Segall, C. Pickard, P. Hasnip, M. Probert, K. Refson and M. Payne, *Z. Kristallogr. Cryst. Mater.* **220**, 567-570 (2005).
4. G. Kresse and J. Furthmüller, *Phys. Rev. B* **54**, 11169 (1996).
5. B. Li, Z. Wan, C. Wang, P. Chen, B. Huang, X. Cheng, Q. Qian, J. Li, Z. Zhang, G. Sun, B. Zhao, H. Ma, R. Wu, Z. Wei, Y. Liu, L. Liao, Y. Yu, Y. Huang, X. Xu, X. Duan, W. Ji and X. Duan, *Nat. Mater.* **20**, 818-825 (2021).
6. M. McGuire, G. Clark, S. KC, W. Chance, G. Jellison, Jr., V. Cooper, X. Xu and B. Sales, *Phys. Rev. Mater.* **1**, 014001 (2017).
7. Z. Zhang, J. Shang, C. Jiang, A. Rasmita, W. Gao and T. Yu, *Nano. Lett.* **19**, 3138-3142 (2019).
8. I. Tsubokawa, *J. Phys. Soc. Jpn.* **15**, 1664 (1960).
9. B. Huang, G. Clark, E. Navarro-Moratalla, D. Klein, R. Cheng, K. Seyler, D. Zhong, E. Schmidgall, M. McGuire, D. Cobden, W. Yao, D. Xiao, P. Jarillo-Herrero and X. Xu, *Nature* **546**, 270 (2017).
10. C. Zhang, L. Wang, Y. Gu, X. Zhang, X. Xia, S. Jiang, L. Huang, Y. Fu, C. Liu, J. Lin, X. Zou, H. Su, J. Mei and J. Dai, *Nanoscale.* **14**, 5851-5858 (2022).
11. C. Gong, L. Li, Z. Li, H. Ji, A. Stern, Y. Xia, T. Cao, W. Bao, C. Wang, Y. Wang, Z. Qiu, R. Cava, S. Louie, J. Xia and X. Zhang, *Nature* **546**, 265 (2017).
12. Y. Sun, Q. Tan, X. Liu, Y. Gao and J. Zhang, *J. Phys. Chem. Lett.* **10**, 3087-3093 (2019).
13. A. Wiedenmann and J. Rossat-Mignod, *Sol. Stat. Comm.* **40**, 1067-1072 (1981).
14. X. Zhang, Q. Lu, W. Liu, W. Niu, J. Sun, J. Cook, M. Vaninger, P. Miceli, D. Singh, S. Lian, T. Chang, X. He, J. Du, L. He, R. Zhang, G. Bian and Y. Xu, *Nat. Comm.* **12**, 2492 (2021).
15. X. Sun, E. Feng, Y. Su, K. Nemkovski, O. Petrscic and T. Brückel, *J. Phys.: Conf. Ser.* **862**, 012027 (2017).
16. W. Roth, *Phys. Rev.* **110**, 1333 (1958).
17. D. Huffman and R. Wild, *Phys. Rev.* **148**, 526 (1966).
18. L. Maxwell and T. McGuire, *Rev. Mod. Phys.* **25**, 279 (1953).
19. K. Kempter and E. Bayer, *IEEE Trans. Mag.* **12**, 62-65 (1976).
20. B. Chamberland, Critical Reviews in Solid State and Materials Sciences **7**, 1-31 (1977).
21. T. Iino, T. Moriyama, H. Iwaki, H. Aono, Y. Shiratsuchi and T. Ono, *App. Phys. Lett.* **114**, 022402 (2019).